\documentclass[aps,pre,preprint,onecolumn,citeautoscript,superscriptaddress,
eqsecnum]{revtex4-1}  
\usepackage{amsmath,amssymb,bm} 
\usepackage{graphicx}  
\usepackage[tight]{subfigure}    
\usepackage{color} 
\usepackage[papersize={8.5in,11in}]{geometry}
\usepackage[colorlinks=true]{hyperref}  
\hypersetup{
    bookmarks=true,         
    unicode=false,          
    pdftoolbar=true,        
    pdfmenubar=true,        
    pdffitwindow=false,     
    pdfstartview={FitH},    
    pdftitle={The novel metallic states of the cuprates:\\ topological Fermi liquids and strange metals},    
    pdfauthor={Subir Sachdev, Debanjan Chowdhury},     
    pdfsubject={},   
    pdfcreator={Creator},   
    pdfproducer={Producer}, 
    pdfkeywords={} {} {}, 
    pdfnewwindow=true,      
    colorlinks=true,       
    linkcolor=magenta, 
    citecolor=blue,        
    filecolor=magenta,      
    urlcolor=blue           
} 

\usepackage{amsfonts}
\usepackage{upgreek}
\usepackage{slashed}
\usepackage{latexsym}

\usepackage{todonotes}

\newcommand{\beq}{\begin{equation}}
\newcommand{\eeq}{\end{equation}}
\def\bea{\begin{eqnarray}}
\def\eea{\end{eqnarray}}
\newcommand{\nn}{\nonumber \\}

\begin{document}

\noindent
\vspace{-0.475cm}
\hspace{-0.3cm} \includegraphics[height=0.5cm]{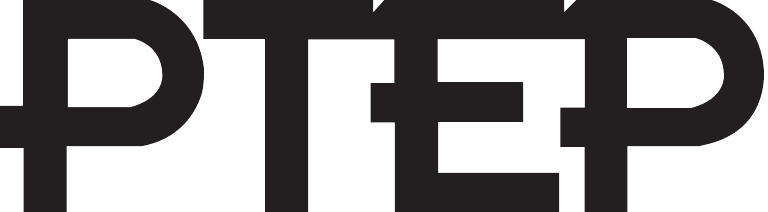} \hfill
{Progress of Theoretical and Experimental Physics} \\
\rule{\textwidth}{0.6pt}
\vspace{0.6cm}

\preprint{\href{http://doi.org/10.1093/ptep/ptw110}{PTEP (2016) 12C102}; \href{http://arXiv.org/abs/1605.03579}{arXiv:1605.03579}}
\title{The novel metallic states of the cuprates:\\ Fermi liquids with topological order, and strange metals}

\author{Subir Sachdev}

\affiliation{Department of Physics, Harvard University, Cambridge MA 02138, USA}
\affiliation{Perimeter Institute for Theoretical Physics, Waterloo, Ontario, Canada N2L 2Y5}

\author{Debanjan Chowdhury}

\affiliation{Department of Physics, Harvard University, Cambridge MA 02138, USA}

\date{May 11, 2016\\
\vspace{0.4in}}

\begin{abstract}%
This article is based on a talk by S.S. at the Nambu Memorial Symposium at the University of Chicago.
We review ideas on the nature of the metallic states of the hole-doped cuprate high temperature
superconductors, with an emphasis on the connections between the Luttinger theorem for the size of the Fermi surface,
topological quantum field theories (TQFTs), and critical theories involving changes in the size of the Fermi surface.
We begin with the derivation of the Luttinger theorem for a Fermi liquid, using momentum balance during a process of flux-insertion
in a lattice electronic model with toroidal boundary conditions. We then review the TQFT of 
the $\mathbb{Z}_2$ spin liquid, and demonstrate its compatibility with the toroidal momentum balance argument. This discussion
leads naturally to a simple construction of Fermi liquid-like states with topological order: the fractionalized Fermi liquid (FL*)
and the algebraic charge liquid (ACL). We present arguments for a description of the pseudogap metal of the cuprates
using $\mathbb{Z}_2$-FL* or $\mathbb{Z}_2$-ACL states with Ising-nematic order. 
These pseudogap metal states are also described as Higgs phases
of a SU(2) gauge theory. The Higgs field represents local antiferromagnetism, but the Higgs-condensed phase does not
have long-range antiferromagnetic order: the magnitude of the Higgs field
determines the pseudogap, the reconstruction of the Fermi surface, and the Ising-nematic order. 
Finally, we discuss the route to the large Fermi surface
Fermi liquid via the critical point where the Higgs condensate and Ising nematic order vanish, 
and the application of Higgs criticality to the strange metal.\\
\\
\end{abstract}


\maketitle

\section{Introduction}
\label{sec:intro}

Nambu's early papers \cite{Nambu60,NJL1,NJL2} laid down the close connection between 
fundamental questions in superconductivity and high energy physics. These connections have continued to flourish to the
present day, to the mutual benefit of both fields. In Ref.~\cite{Nambu60}, Nambu clarified the manner in which gauge-invariance
was maintained in the BCS theory of the Meissner effect of superconductivity, and this paved the way for the proposal of the Higgs-Anderson mechanism.
The subsequent papers \cite{NJL1,NJL2} treated the BCS theory in a slightly different manner: 
it was viewed as a theory with a {\em global\/} U(1) symmetry, rather than with the U(1) gauge invariance of Maxwell electromagnetism.
The breaking of the global U(1) symmetry led to the appearance of Nambu-Goldstone bosons, and this inspired ideas on chiral symmetry breaking
in nuclear physics. 
These global and gauge perspectives on electromagnetism turn out to be closely related because the electromagnetic theory is weakly coupled,
but it is important to keep the distinction in mind.

In the present article, in the hopes of continuing the tradition pioneered by Nambu, we will review recently discussed connections between
the {\em high\/} temperature superconductors and gauge theories. The gauge theories will all involve strongly-coupled {\em emergent\/} gauge fields,
while the U(1) gauge invariance of electromagnetism will be treated as a global symmetry. In this context, the emergent gauge fields do not reflect any
underlying symmetry of the Hamiltonian, but are a manifestation of the long-range quantum entanglement of the states under consideration.
We will illustrate how emergent gauge fields are powerful tools for deducing the physical properties of entangled many-body quantum states,
and for connecting theories to experimental observations.

\begin{figure}
\begin{center}
\includegraphics[height=7.5cm]{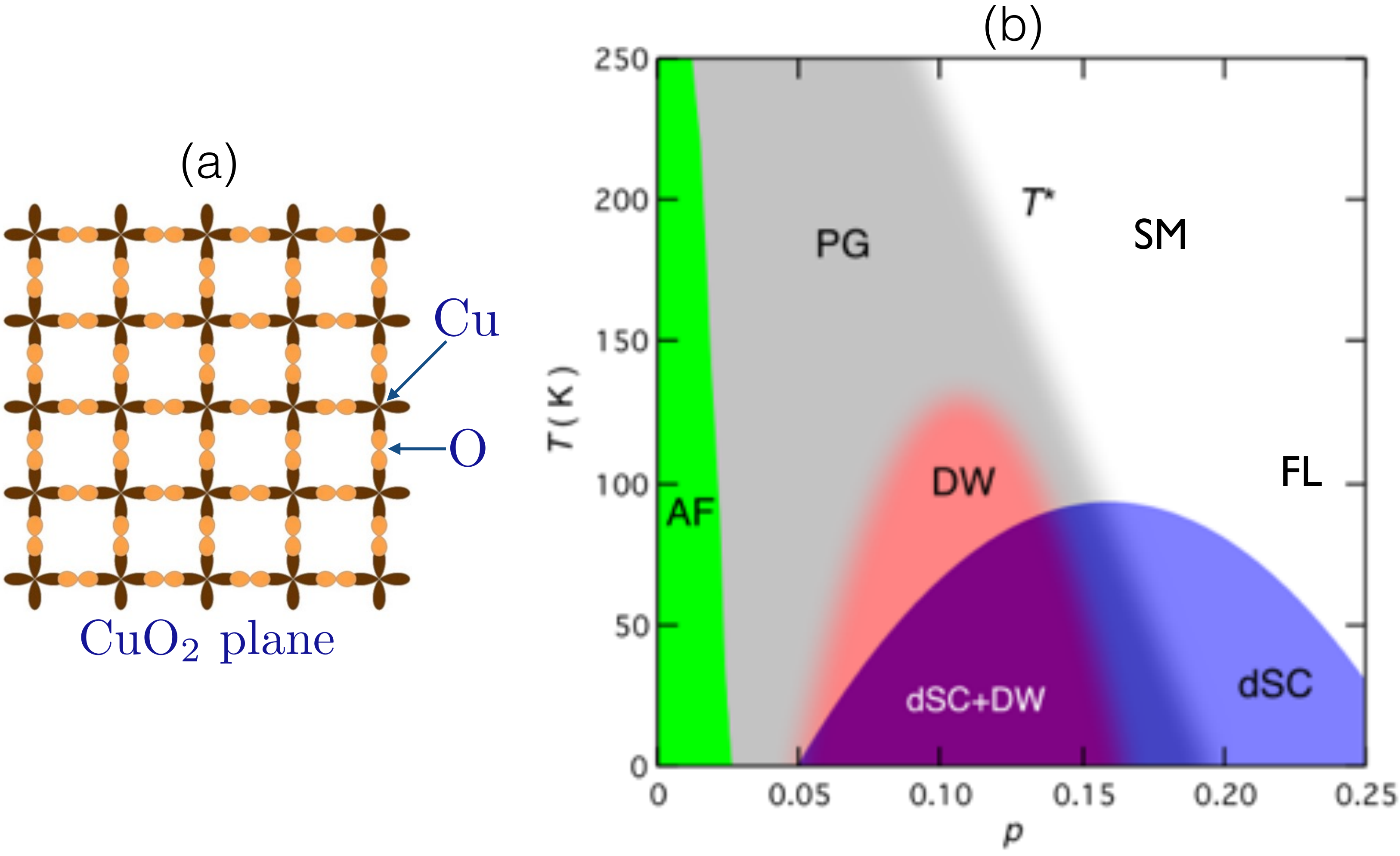}
\end{center}
\caption{(a) The square lattice of Cu and O atoms found in every copper-based high temperature superconductor.
(b) A schematic phase diagram of the YBCO superconductors as a function of the hole density $p$
and the temperature $T$ adapted from Ref.~\cite{MHH15a}. 
The phases are discussed in the text: AF--insulating antiferromagnet, PG--pseudogap, 
DW--density wave, dSC--$d$-wave superconductor, SM--strange metal, FL--Fermi liquid. The critical temperature for superconductivity
is $T_c$, and $T^\ast$ is the boundary of the pseudogap regime.}
\label{fig:phasediag}
\end{figure}

\begin{figure}
\begin{center}
\includegraphics[height=5cm]{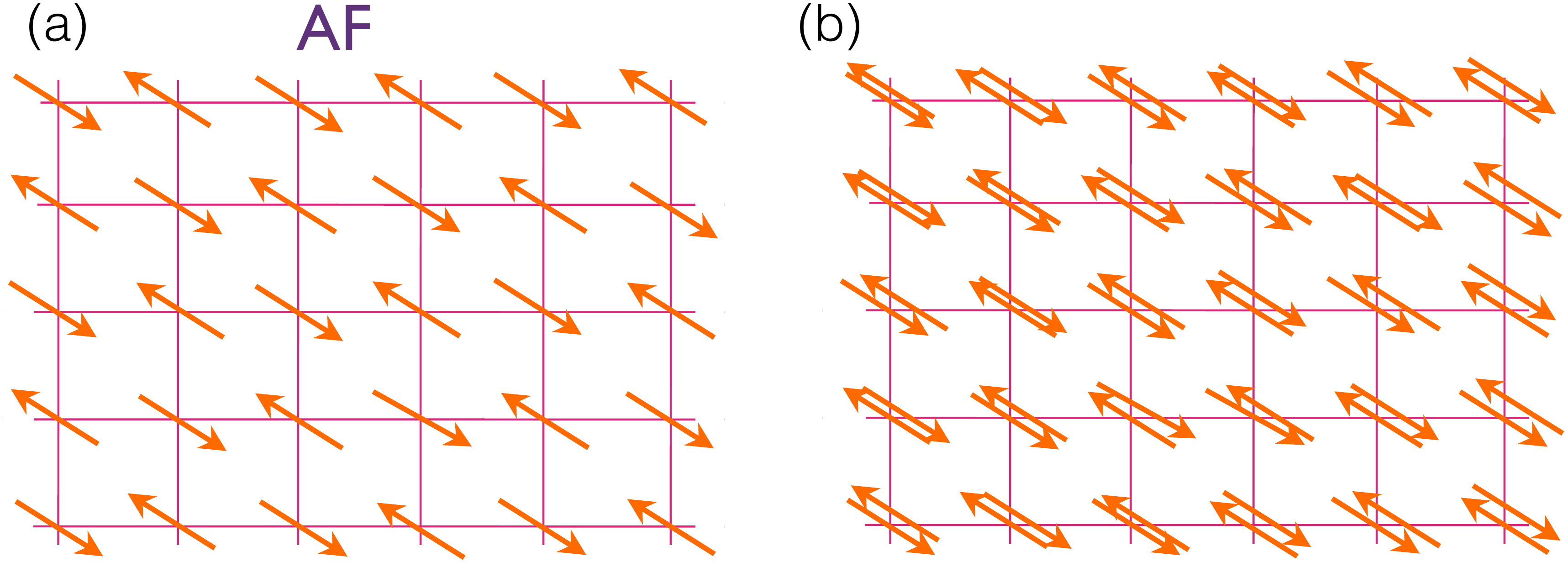}
\end{center}
\caption{(a) The insulating AF state at hole density $p=0$. (b) The band insulator with 2 electrons per site.}
\label{fig:af}
\end{figure}

In Fig.~\ref{fig:phasediag}a, we show the quasi-two dimensional layers of CuO$_2$. 
For the purposes of this article, we can regard the O $p$ orbitals as filled with pairs
of electrons and inert. Only one of the Cu orbitals is active, and in a parent insulating compound, this orbital has
a density of exactly one electron per site. The rest of this article will consider the physical properties of this Cu orbital
residing on the vertices of a square lattice. Fig.~\ref{fig:phasediag}b shows a schematic phase diagram of the hole-doped copper oxide superconductors.
The AF state in Fig.~\ref{fig:phasediag}b is the antiferromagnet shown in Fig.~\ref{fig:af}a,
in which there is one electron on each Cu orbital, and their spins are polarized in a checkerboard pattern. This state is referred to as a Mott
insulator, because it is primarily the Coulomb repulsion which prevents the electrons from becoming mobile.
This AF insulator should be contrasted from the band insulator with 2 electrons per Cu site, which is shown in Fig.~\ref{fig:af}b; the latter
state is an insulator even for non-interacting electrons because all electron motion is impeded by the Pauli exclusion principle.

\begin{figure}
\begin{center}
\includegraphics[height=5.75cm]{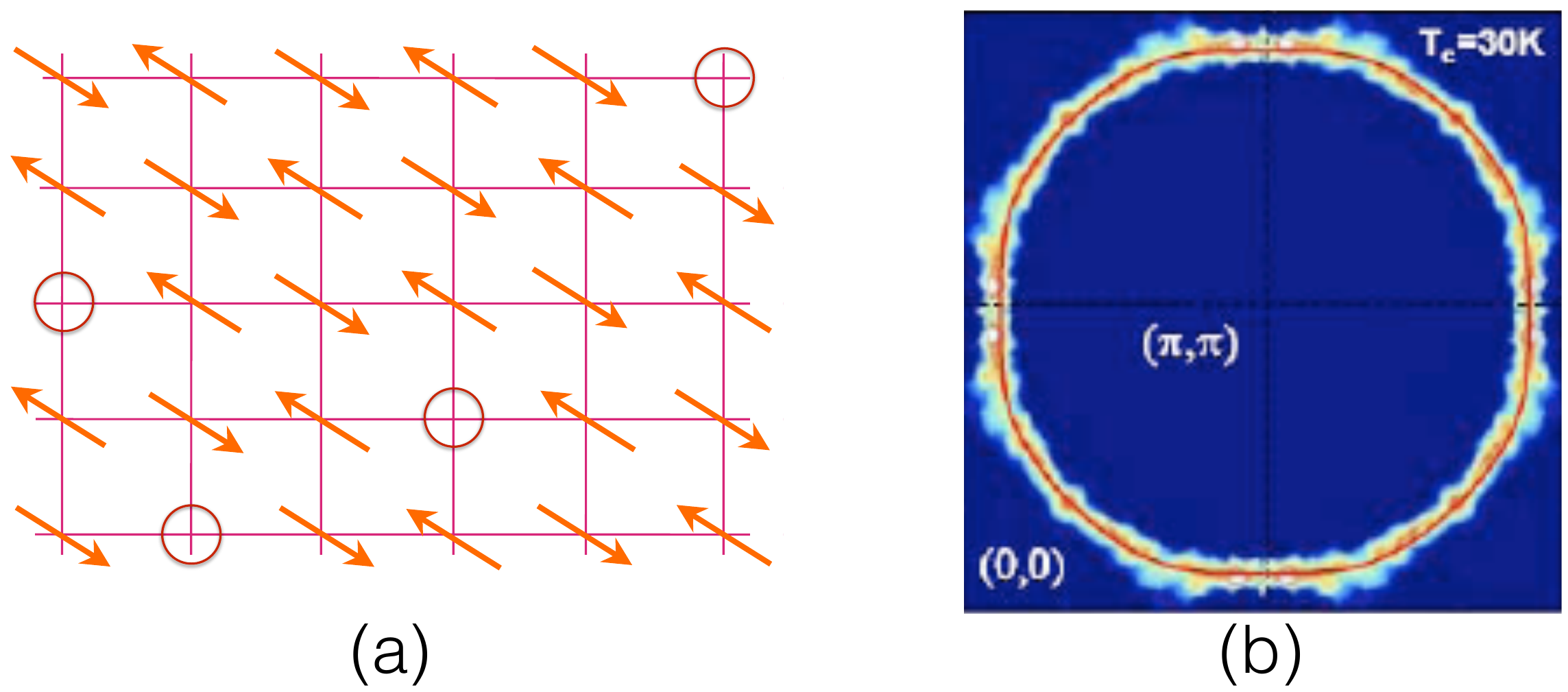}
\end{center}
\caption{(a) State obtained after removing electrons with density $p$ from the AF state in Fig.~\ref{fig:af}a.
Relative to the fully-filled state with 2 electrons per site in Fig.~\ref{fig:af}b, this state has a density of holes equal to $1+p$.
(b) Photoemission results from Ref.~\cite{Dama05} showing a Fermi surface of size $1+p$ in the FL region of Fig.~\ref{fig:phasediag}b. This is the
Fermi surface size expected by the Luttinger theorem for a Fermi liquid without AF order or other broken symmetry.}
\label{fig:fl}
\end{figure}

The rich phases of the cuprates appear when we remove a density of $p$ electrons from the AF state, as illustrated in Fig.~\ref{fig:fl}a.
It is important to note that relative to the band insulator in Fig.~\ref{fig:af}b, the state in Fig.~\ref{fig:fl}a has a density of $1+p$
holes.
So if we described the ground state at this density by adiabatic continuity from a free electron ground state, 
the Luttinger theorem states that we should obtain a 
metal with a Fermi surface of size equivalent to $1+p$ holes. This turns out to be precisely the case in the larger $p$ region labeled FL 
(for Fermi liquid) in Fig.~\ref{fig:phasediag}b. The corresponding `large' Fermi surface observed in photoemission experiments
is shown in Fig.~\ref{fig:fl}b.

The focus of this article will be on the metallic phases in Fig.~\ref{fig:phasediag}b, labeled by PG, SM, and FL.
Of these, only the FL appears to be well understood as a conventional Fermi liquid. The traditional proof of the Luttinger theorem is given in terms of conventional diagrammatic and Ward identity arguments.
However, it was argued more recently by Oshikawa \cite{MO00} 
that the Luttinger theorem has a topological character, and a proof can be given
using a momentum balance 
argument that follows the many-electron wavefunction on a torus geometry in the presence of a flux penetrating one of the cycles
of the torus. We will review this argument in Section~\ref{sec:fl}. The subsequent Section~\ref{sec:sl} will turn to spin liquid states of the 
insulator at $p=0$: these states are described at low energies by a topological quantum field theory (TQFT). 
We will describe key characteristics of the TQFT which enable the spin liquid to also satisfy the momentum balance constraints of
Section~\ref{sec:fl}.

We will describe a model for the 
pseudogap (PG) metal as a $\mathbb{Z}_2$-FL* state (and the related $\mathbb{Z}_2$-ACL state) 
in Section~\ref{sec:fls}, along with its connections to 
recent experimental observations. The strange metal (SM) appears to be a metal without quasiparticle excitations,
and we will discuss candidate critical field theories for such a state in Section~\ref{sec:sm}.

A small part of 
the discussion in Sections~\ref{sec:sl} and~\ref{sec:fls} overlaps with a separate, less technical, recent article by one of us \cite{SSroyal}.

\section{Momentum balance on the torus and the Luttinger theorem}
\label{sec:fl}

Consider an arbitrary quantum system, of bosons or fermions, 
defined on (say) a square lattice of unit lattice spacing, and placed on a torus.
The size of the lattice is $L_x \times L_y$, and we impose periodic boundary conditions. Assume the system has a global U(1)
symmetry, and all the local operators carry integer U(1) charges. 
Pick an eigenstate of the Hamiltonian (usually the ground state) $\left| G \right\rangle$.
Because of the translational symmetry, this state will obey
\beq
\hat{T}_x \left| G \right\rangle = e^{i P_x} \left| G \right\rangle,
\eeq
where $\hat{T}_x$ is the translational operator by one lattice spacing along the $x$ direction, 
and $P_x$ is the momentum of the state $\left| G \right\rangle$.
Note that $P_x$ is only defined modulo $ 2\pi$. The state $\left| G \right\rangle$ will also have definite total U(1) charge, which we denote
by the integer $N$. 

\begin{figure}
\begin{center}
\includegraphics[height=3.5cm]{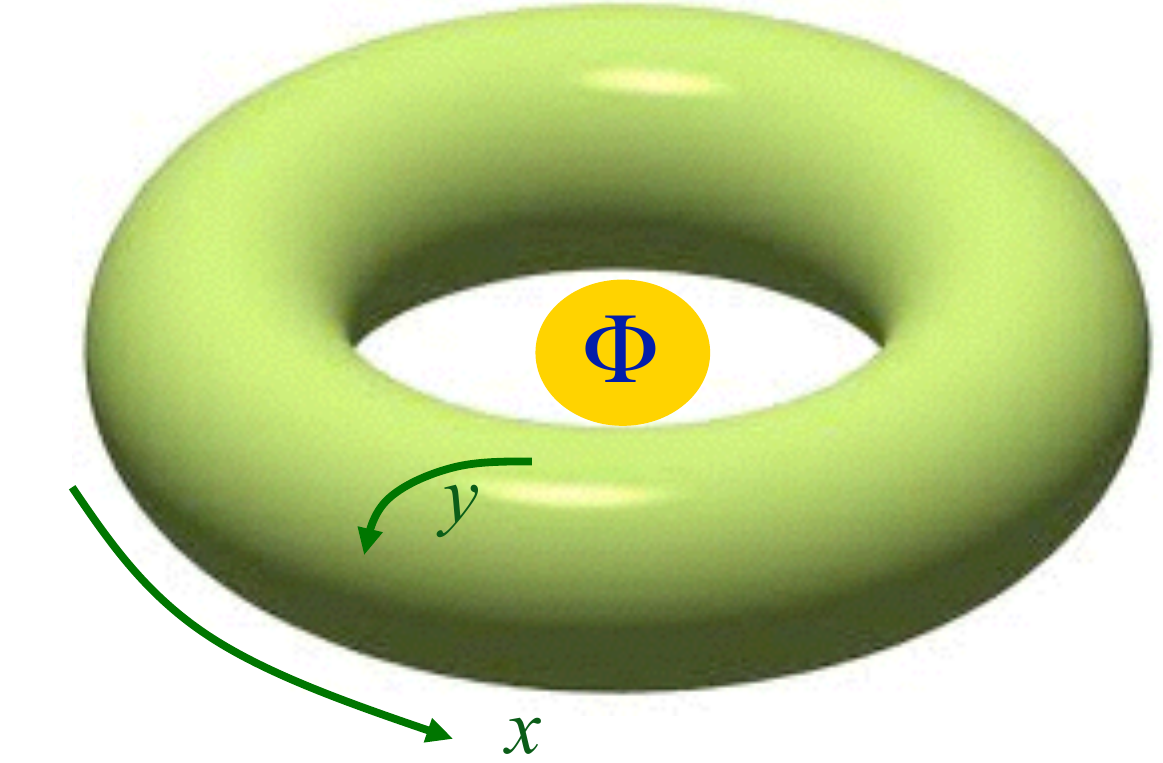}
\end{center}
\caption{Torus geometry with a flux quantum inserted.}
\label{fig:torus}
\end{figure}

Now we gauge the global U(1) symmetry, and insert one flux quantum (with flux $2 \pi$) through one of the cycles of the torus
(see Fig.~\ref{fig:torus}).
After the flux insertion, the Hamiltonian is gauge equivalent to the Hamiltonian without the flux. So we gauge transform to the original
Hamiltonian; the new state of the system, $\left| G' \right\rangle$ will not, in general, be the same as the original state $\left| G \right\rangle$.
Indeed, its momentum $P_x'$ will differ from $P_x$ by $\Delta P_x$ with
\beq
\Delta P_x = \frac{2 \pi}{L_x} N \, (\mbox{mod} \, 2 \pi).
\label{deltaP}
\eeq
A general proof of (\ref{deltaP}) can be found in Refs.~\cite{MO00,TSMVSS04,APAV04}. But we can easily deduce
the result by first considering the case of non-interacting particles. Then, an elementary argument shows that each particle 
picks up momentum $2 \pi/L_x$ from the flux insertion, and so (\ref{deltaP}) is clearly valid. Now turn on the interactions: these cannot
change the total momentum, which is conserved (modulo $2 \pi$) both by the interactions and the flux insertion; so (\ref{deltaP}) 
applies also in the presence of interactions.

So far, we have been quite general, and not specified anything about the many-body system, apart from
its translational invariance and global U(1) symmetry. In the subsequent discussion, we will make further assumptions
about the nature of the ground state and low-lying excitations, and compute $\Delta P$ by other methods. Equating such a result
to (\ref{deltaP}) will then lead to important constraints on the allowed structure of the many-body ground state.

In the present section, following Oshikawa \cite{MO00}, we assume the ground state is a Fermi liquid.
So its only low-lying excitations are fermionic quasiparticles around the Fermi surface. For our subsequent discussion,
it is important to also include the electron spin index, $\alpha = \uparrow, \downarrow$, 
and so we will have a Fermi liquid with 2 global U(1) symmetries, associated
respectively with the conservation of electron number and the $z$-component of the total spin, $S_z$. Consequently, there will  be two Luttinger
theorems, one for each global U(1) symmetry. The action for the fermionic quasiparticles, $c_{{\bm k}\alpha}$, with dispersion 
$\varepsilon ({\bm k})$ is
\beq
\mathcal{S}_{FL} = \int d\tau \int \frac{d^2 k}{4 \pi^2} \sum_{\alpha = \pm 1} 
c_{{\bm k} \alpha}^\dagger \left( \frac{\partial}{\partial \tau} 
- \frac{i}{2} \alpha A_\tau^s - i A_\tau^e  + \varepsilon ({\bm k} - \alpha {\bm A}^s /2 - {\bm A}^e) \right) c_{{\bm k} \alpha} \, ,
\label{SFL}
\eeq
where $\tau$ is imaginary time. The Fermi surface is defined by $\varepsilon ({\bm k}) = 0$, and $\mathcal{S}_{FL}$ only applies for ${\bm k}$
near the Fermi surface, although we have (for notational convenience) written it in terms of an integral over all ${\bm k}$.
We have also coupled the quasiparticles to 2 probe gauge fields $A_\mu^e = (A_\tau^e, {\bm A}^e)$ and 
$A_\mu^s = (A_\tau^s, {\bm A}^s)$ which couple to the 2 conserved U(1) currents associated, respectively, with the conservation of electron
number and $S_z$.

We place the Fermi liquid on a torus, and insert a $2 \pi$ flux of a gauge field that couples only to the up-spin electrons.
So we choose $A_\mu^s = 2 A_\mu^e \equiv A_\mu$. Then the general momentum balance in (\ref{deltaP}) requires that
\beq
\Delta P_x = \frac{2 \pi}{L_x} N_\uparrow \, (\mbox{mod} \, 2 \pi) = \frac{2 \pi}{L_x} \frac{N}{2} \, (\mbox{mod} \, 2 \pi),
\label{deltaP2}
\eeq
where we assume equal numbers of up and down spin electrons $N_\uparrow = N_\downarrow = N/2$.
Now let us determine $\Delta P_x$ by using the description of the quasiparticles described by $\mathcal{S}_{FL}$. 
\begin{figure}
\begin{center}
\includegraphics[height=5cm]{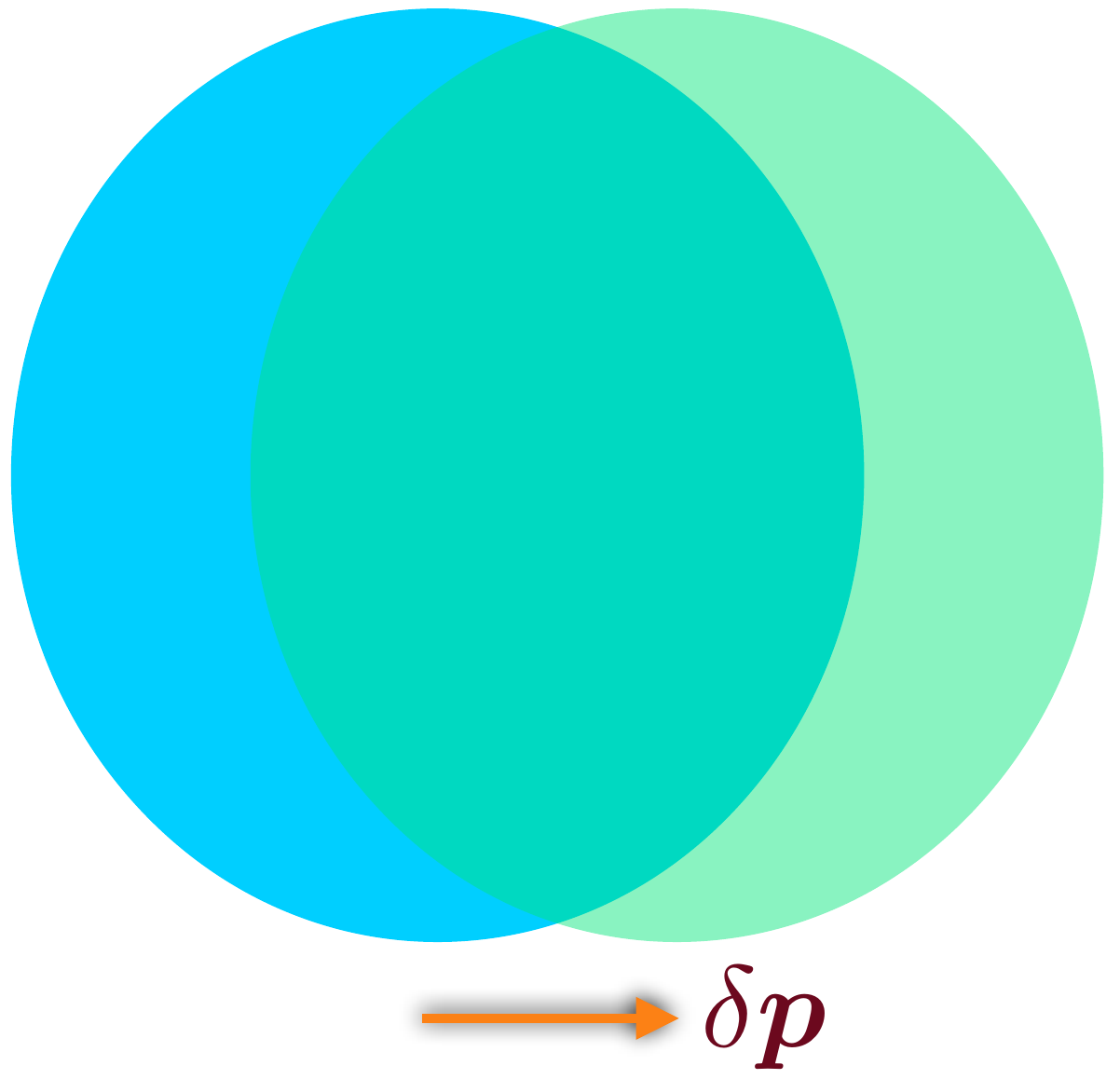}
\end{center}
\caption{Response of a Fermi liquid to flux insertion. Each quasiparticle near the Fermi surface acquires a momentum shift
${\delta {\bm p}} = (\delta p_x , 0)$. The total change in momentum is equal to the difference in the total momenta between the
blue and green regions. This equality assumes quasiparticles exist at all momenta, but this is permissible because the net
contribution arises only from the regions near the Fermi surface, where the quasiparticles do exist.}
\label{fig:fsmove}
\end{figure}
As illustrated in Fig.~\ref{fig:fsmove}, each quasiparticle near the Fermi surface will behave like a free fermion, and have its 
momentum shifted by $\delta p_x = 2 \pi/L_x$. We add up the contributions of all the quasiparticles by integrating in the 
vicinity of the Fermi surface. After using the divergence theorem, or pictorially by the sketch in Fig.~\ref{fig:fsmove}, we can convert
the integral to a volume integral inside the Fermi surface \cite{MO00,APAV04}, and so show
\beq
\Delta P_x = \frac{2 \pi}{L_x} \left( L_x L_y \frac{V_{FS}}{4 \pi^2} \right) \, (\mbox{mod} \, 2 \pi),
\label{Vfs}
\eeq
where $V_{FS}$ is the momentum space area enclosed by the Fermi surface; the factor within the brackets on the right-hand-side equals
the number of momentum space points inside the Fermi surface. Note that the entire contribution to the right-hand-side of (\ref{Vfs}) comes
from the vicinity of the Fermi surface where the quasiparticles are well-defined; we have merely used a mathematical identity to convert
the result to the volume, and we are not assuming the existence of quasiparticles far from the Fermi surface.

Now we use (\ref{deltaP2}) and (\ref{Vfs}), along with the corresponding
expressions for flux inserted in the other cycle of the torus, to deduce the Luttinger theorem. The complete argument requires careful
attention to the $(\mbox{mod} \, 2 \pi)$ factors using situations where $L_x$ and $L_y$ are mutually prime integers \cite{MO00,APAV04}.
But ultimately, naively equating (\ref{deltaP2}) and (\ref{Vfs}) gives the correct result
\beq
\frac{V_{FS}}{2 \pi^2} = \frac{N}{L_x L_y} \, (\mbox{mod} \, 2) = (1+p) \, (\mbox{mod} \, 2).
\label{LT}
\eeq
In the final step, we have applied the Luttinger theorem to the holes in the cuprates, with a density of holes of $(1+p)$
relative to the filled band insulator in Fig.~\ref{fig:af}b. The expression (\ref{LT}) is experimentally verified in the FL region
in Fig.~\ref{fig:fl}b.

\section{Topological quantum field theory of the $\mathbb{Z}_2$ spin liquid}
\label{sec:sl}

\begin{figure}
\begin{center}
\includegraphics[height=10cm]{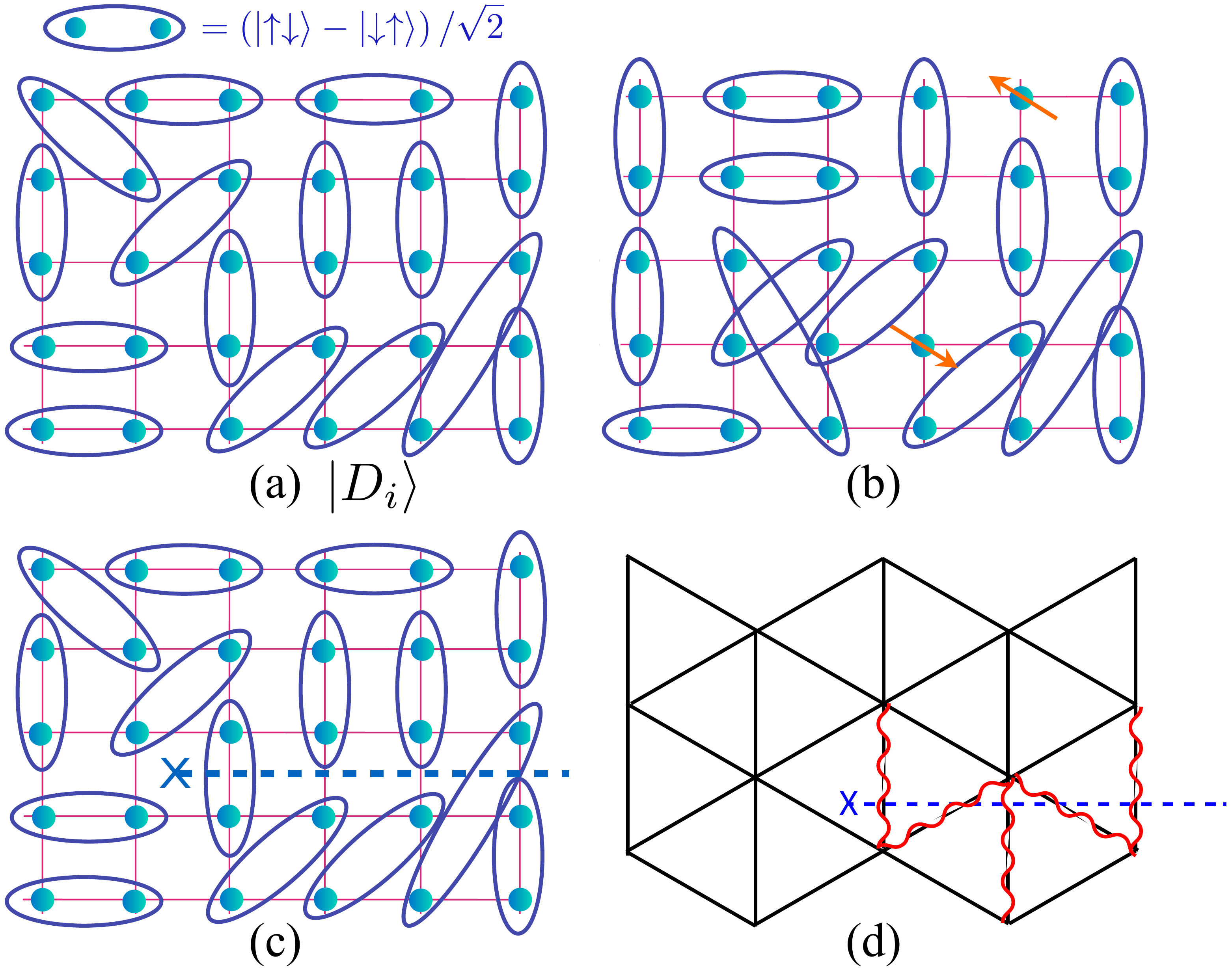}
\end{center}
\caption{(a) Illustration of a component, $\left| D_i \right\rangle$, of the RVB wavefunction in (\ref{eq:rvb}).
(b) A pair of $S=1/2$ spinon excitations. (c) The vison excitation of the $\mathbb{Z}_2$ spin liquid. In terms of (\ref{eq:rvb}), 
the co-efficients $c_i$ are modified so that each singlet bond crossing the `branch-cut' (dashed line) picks up a factor of $-1$.
A similar modification applies to (\ref{spinl}), and is described in the text. In the TQFT, the branch-cut is represented
by (\ref{WL}). (d) A vison on the triangular lattice for the case
of $Q_{ij}$ and $P_{ij}$ non-zero only between nearest-neighbor sites: the wavy lines indicate the $Q_{ij}$ and $P_{ij}$ with
a change in their sign in the presence of a vison.
}
\label{fig:rvb}
\end{figure}

We now return to the insulator at $p=0$. In Fig.~\ref{fig:phasediag}b, the insulator breaks translational and spin rotation symmetries
in the AF state shown in Fig.~\ref{fig:af}a. However, as AF order disappears at rather small values of $p$, it is useful
to begin the analysis of doped states by 
examining insulating states at $p=0$ which preserve both translation and spin rotation symmetries. An example of such a state is the 
`resonating valence bond' (RVB) insulator \cite{Pauling49,Anderson73,GBPWA88,DRSK88}, illustrated in Fig.~\ref{fig:rvb}a. A trial wavefunction
for the RVB state takes the form
\beq
\left| \Psi \right\rangle = \sum_i c_i \left| D_i \right \rangle \label{eq:rvb}
\eeq
where $i$ extends over all possible pairings of electrons on nearby sites, and a state $\left| D_i \right \rangle$ associated with
one such pairing is shown in Fig.~\ref{fig:rvb}a. 
Note that the electrons in a valence bond need not be nearest-neighbors.
Each $\left| D_i \right \rangle$ is a spin singlet, and so spin rotation invariance is preserved.
We also assume that the $c_i$ respect the translational
and other symmetries of the square lattice.

A theory for a stable RVB state with time-reversal symmetry and a gap to all excitations first appeared in Refs.~\cite{NRSS91,XGW91,SSkagome},
which described a state now called a $\mathbb{Z}_2$ spin liquid. It is helpful to describe the structure of the $\mathbb{Z}_2$ spin liquid
in terms of a mean-field ansatz. 
We write the spin operators on each site, ${S}_{i\ell}$ ($\ell = x,y,z$), 
in terms of Schwinger bosons $b_{i \alpha}$ ($\alpha = \uparrow, \downarrow$)
\beq
{S}_{i\ell} = \frac{1}{2} b_{i \alpha}^\dagger {\sigma}^\ell_{\alpha\beta} b_{i \beta}, \label{Sbb}
\eeq
where ${\sigma}^\ell$ are the Pauli matrices, and the bosons obey the local constraint
\beq
\sum_\alpha b_{i \alpha}^\dagger b_{i \alpha} = 2S \label{const}
\eeq
on every site $i$. Here we are primarily interested in the case of spin $S=1/2$, but it is useful to also consider the case of general $S$.
Schwinger fermions can also be used instead, but the description of the $S>1/2$ cases is more cumbersome with them.
The $\mathbb{Z}_2$ spin liquid is described by an effective Schwinger boson Hamiltonian \cite{AA88,NRSS91}
\beq
\mathcal{H}_b =  - \sum_{i < j} \left[ P_{ij} b_{i \alpha}^\dagger b_{j \alpha} +  Q_{ij} \varepsilon_{\alpha\beta} b_{i \alpha}^\dagger b_{j \beta}^\dagger + \mbox{H.c.} \right] + \lambda \sum_i b_{i \alpha}^\dagger
b_{i \alpha}, \label{Hb}
\eeq
where $\varepsilon_{\alpha\beta}$ is the antisymmetric unit tensor, $\lambda$ is chosen to satisfy the constraint in Eq.~(\ref{const}) on average, 
and the $Q_{ij}=-Q_{ji}$ and $P_{ij} = P_{ji}^\ast$ are a set of variational parameters chosen to optimize the energy of the spin liquid state. Generally, the $Q_{ij}$ and $P_{ij}$ are
chosen to be non-zero only between nearby sites, and  the `$\mathbb{Z}_2$' character of the spin liquid requires that the links with non-zero $Q_{ij}$
can form closed loops with an odd number of links: the Schwinger boson parameterization (\ref{Sbb}) is invariant under the 
U(1) gauge transformation, $b_{i \alpha} \rightarrow e^{i \phi_i} b_{i \alpha}$, and odd loops imply that the U(1) is 
Higgsed down to a $\mathbb{Z}_2$ gauge theory \cite{NRSS91,RJSS91,XGW91,SSkagome,MVSS99}.
This Hamiltonian yields a mean-field wavefunction for the spin liquid
\beq
\left|\Psi \right\rangle = \mathcal{P}_{2S} \exp \left( \sum_{i<j} f_{ij} \, \varepsilon_{\alpha\beta} b_{i \alpha}^\dagger b_{j \beta}^\dagger
\right) |0 \rangle, \label{spinl}
\eeq
where $|0 \rangle$ is the boson vaccum, $\mathcal{P}_{2S}$ is a projection operator which selects only states which obey Eq.~(\ref{const}), 
and the boson pair wavefunction $f_{ij}=-f_{ji}$ is determined by diagonalizing Eq.~(\ref{Hb}) by a Bogoliubov transformation.

Moving to the gapped excited states of the $\mathbb{Z}_2$ spin liquid, we find two distinct types of quasiparticles, illustrated in 
Fig.~\ref{fig:rvb}b-d.\\ 
({\em i\/}) A `spinon', shown in Fig.~\ref{fig:rvb}b, has one unpaired spin and so carries spin $S=1/2$; more specifically, the
spinon is the Bogoliubov quasiparticle obtained by diagonalizing $\mathcal{H}_b$ in terms of canonical bosons.\\
({\em ii\/}) The second quasiparticle, the `vison', shown in Fig.~\ref{fig:rvb}c,d, is spinless and it has a more subtle topological character
of a vortex in an Ising-like system (hence its name \cite{TSMPAF00}).
The vison state is the ground state of a Hamiltonian, $\mathcal{H}_b^v$, obtained from $\mathcal{H}$ by mapping
$Q_{ij} \rightarrow Q^v_{ij}$, $P_{ij} \rightarrow P^v_{ij}$; then the vison state $|\Psi^v \rangle$ has a wavefunction as in Eq.~(\ref{spinl}),
but with $f_{ij} \rightarrow f^v_{ij}$. 
Far from the center of the vison, we have $|Q^{v}_{ij}| = |Q_{ij}|$, $|P^{v}_{ij}| = |P_{ij}|$, while closer to the center there
are differences in the magnitudes. However, the key difference is in the signs of the  link variables, as illustrated in Fig.~\ref{fig:rvb}c,d:
there is a `branch-cut' emerging from the vison core along which $\mbox{sgn}(Q^{v}_{ij}) = - \mbox{sgn}(Q_{ij})$ and $\mbox{sgn}(P^{v}_{ij}) = - \mbox{sgn}(P_{ij})$. 
This branch-cut
ensures that the $\mathbb{Z}_2$ magnetic 
flux equals -1 on all loops which encircle the vison core, while other loops do not have non-trivial $\mathbb{Z}_2$ flux. 

\begin{figure}
\begin{center}
\includegraphics[height=8cm]{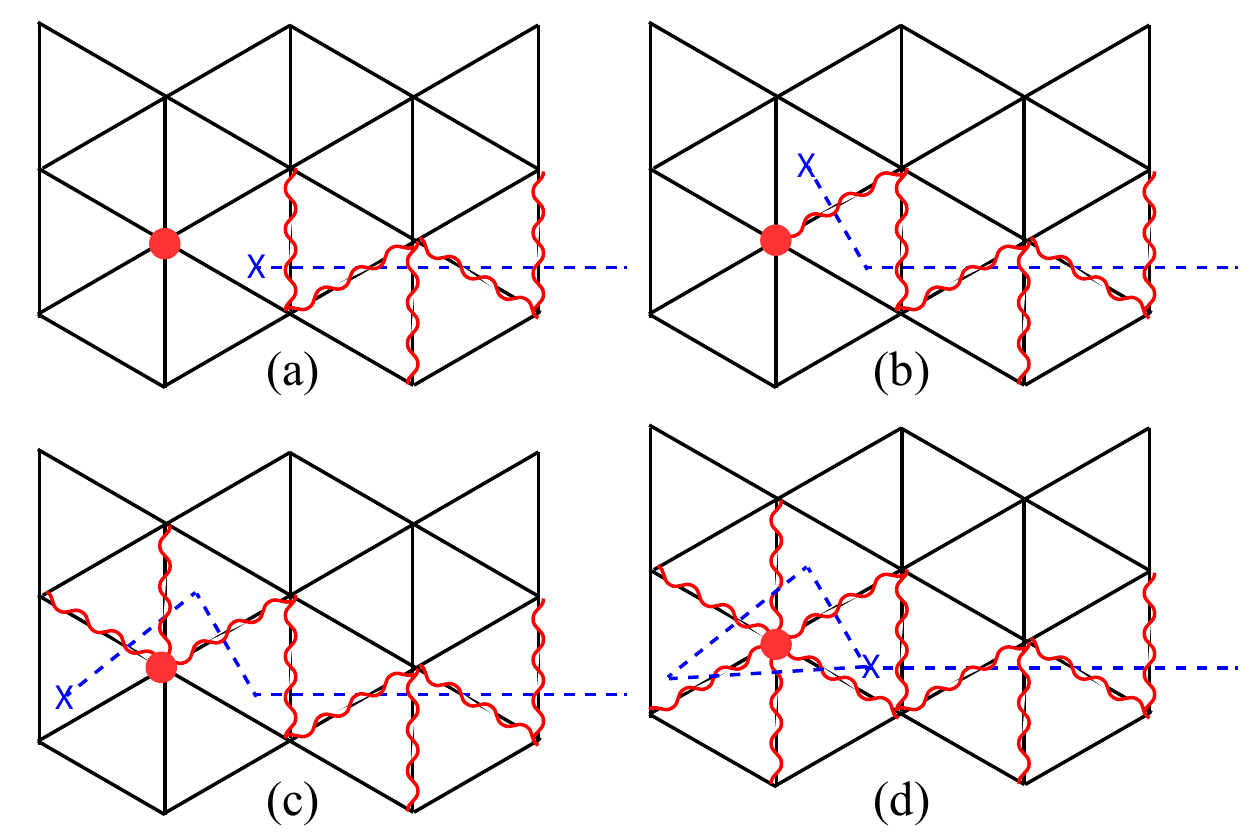}
\end{center}
\caption{Adiabatic motion of a vison (denoted by the X) around a single site of the triangular lattice (denoted by the filled circle).
The initial state is in (a), and the final state is in (d), and these differ by a gauge transformation under which 
$b_{i\alpha} \rightarrow - b_{i\alpha}$
only on the filled circle site.}
\label{fig:visonberry}
\end{figure}

The spinons and visons have two crucial topological properties. \\
({\em i\/}) A spinon and a vison are mutual semions \cite{RC89}. In other words, adiabatically moving a spinon around a vison (or vice versa) 
yields a Berry phase
of $\pi$.  This is evident from the structure of the branch-cut in $Q^v_{ij}$ and $P^v_{ij}$: these $Q^v_{ij}$ and $P^v_{ij}$ are the hopping
amplitudes for the spinon, and they yield an additional phase of $\pi$ (beyond those provided by $P_{ij}$ and $Q_{ij}$) 
every time a spinon crosses the branch cut.\\
({\em ii\/}) A less well-known and distinct property involves the motion of a single vison without any spinons present: 
adiabatic motion of a vison around a single
lattice site yields a Berry phase of $2\pi S$ \cite{RJSS91,MVSS99,TSMPAF00}. This property is illustrated in Fig.~\ref{fig:visonberry}, 
and see Ref.~\cite{YHMPSS11} for a complete computation.
The initial and final states of the adiabatic motion differ by a $\mathbb{Z}_2$ gauge transformation, $b_{i\alpha} \rightarrow - b_{i\alpha}$, only
on the site which has been encircled. From the projection operator $\mathcal{P}_{2S}$ in (\ref{spinl}) we find that
the wavefunction $\left| \Psi \right\rangle$ has picked up a factor of $(-1)^{2 S}$, and this is the only contribution to a gauge-invariant
Berry phase.

The background Berry phase of $2 \pi S$ per site for vison motion 
implies that there are two distinct types of $\mathbb{Z}_2$ spin liquids \cite{RJSS91,MVSS99,TSMPAF00,MSF02,TSOM02,OMTS02}.
As  was first pointed out in Refs.~\cite{RJSS91,MVSS99},
these are `odd $\mathbb{Z}_2$ spin liquids', which are realized in the present model by
half-integer $S$ antiferromagnets, and `even $\mathbb{Z}_2$ spin liquids', realized here by integer $S$ antiferromagnets. 
In the $\mathbb{Z}_2$ gauge theory framework (or the related `toric code' \cite{Kitaev03}), 
there is a unit $\mathbb{Z}_2$ electric
charge on each lattice site of an odd-$\mathbb{Z}_2$ gauge theory.

\subsection{Topological quantum field theory}

All of the above properties of the $\mathbb{Z}_2$ spin liquid can be described elegantly using a topological quantum field theory (TQFT).
The TQFT presentation also highlights the robustness and generality of the structure we have described above.

The TQFT is obtained by implementing the mutual semion statistics between the spinon and the vison using U(1) Chern-Simons gauge
fields. We introduce two `emergent' gauge fields, $a_\mu$ and $b_\mu$. We couple the visons to $a_\mu$ with unit charge.
This implies that the branch-cut emanating from the vison in Fig.~\ref{fig:rvb}c, d is the Wilson line operator
\beq
\exp\left( i \int_{\mathcal{B}}
dx_i a_i \right), \label{WL}
\eeq
taken along the
branch cut $\mathcal{B}$.
We couple the spinons to $b_\mu$, also with  unit charge. 
We also note that the external gauge field $A_\mu^s$ coupling to the 
$z$-component of the spin $S_z$ (see (\ref{SFL})) will also couple to the spinons which carry $S_z = \pm 1/2$.
Then standard methods \cite{FWAZ83} yield the following action for the TQFT (in imaginary time, $\tau$) \cite{Freedman04}
\beq
\mathcal{S}_{CS} = \int d^2 x d \tau \left[ \frac{i}{\pi} \epsilon_{\mu\nu\lambda} a_\mu \partial_\nu b_\lambda + 
\frac{i}{2 \pi} \epsilon_{\mu\nu\lambda} A_\mu^s \partial_\nu a_\lambda \right].
\label{SCS}
\eeq
This theory can be exactly quantized \cite{witten89}, and this yields interesting information on the structure of $\mathbb{Z}_2$ spin
liquids on topologically non-trivial manifolds. On a torus, the only non-trivial gauge-invariant observables are the Wilson loops
around the two cycles of the torus, which we denote by 
\bea
\hat{W}_x = \exp \left( i \oint dx \, a_x \right) \quad &,& \quad \hat{W}_y = \exp \left( i \oint dy \, a_y \right) \nn
\hat{V}_x = \exp \left( i \oint dx \, b_x \right) \quad &,& \quad \hat{V}_y = \exp \left( i \oint dy \, b_y \right). \label{4fold}
\eea
The quantization of (\ref{SCS}) at $A_\mu^s = 0$ is characterized by the operator algebra
\beq
\hat{W}_x \hat{V}_y = - \hat{V}_y \hat{W}_x \quad, \quad \hat{W}_y \hat{V}_x = - \hat{V}_x \hat{W}_y ,
\eeq
and all other combinations of operators commute. This operator algebra is easily realized by 2 independent sets of Pauli matrices.
This implies that the ground state of the $\mathbb{Z}_2$ spin liquid has a 4-fold degeneracy on the torus. This degeneracy
can also be obtained from the trial wavefunctions for the spin liquid \cite{Thouless87,KRS88} 
in Section~\ref{sec:sl}: the degenerate ground states are obtained
by applying the branch-cut around the cycles of the torus, a connection evident from (\ref{WL}) and (\ref{4fold}).

\begin{figure}
\begin{center}
\includegraphics[height=5cm]{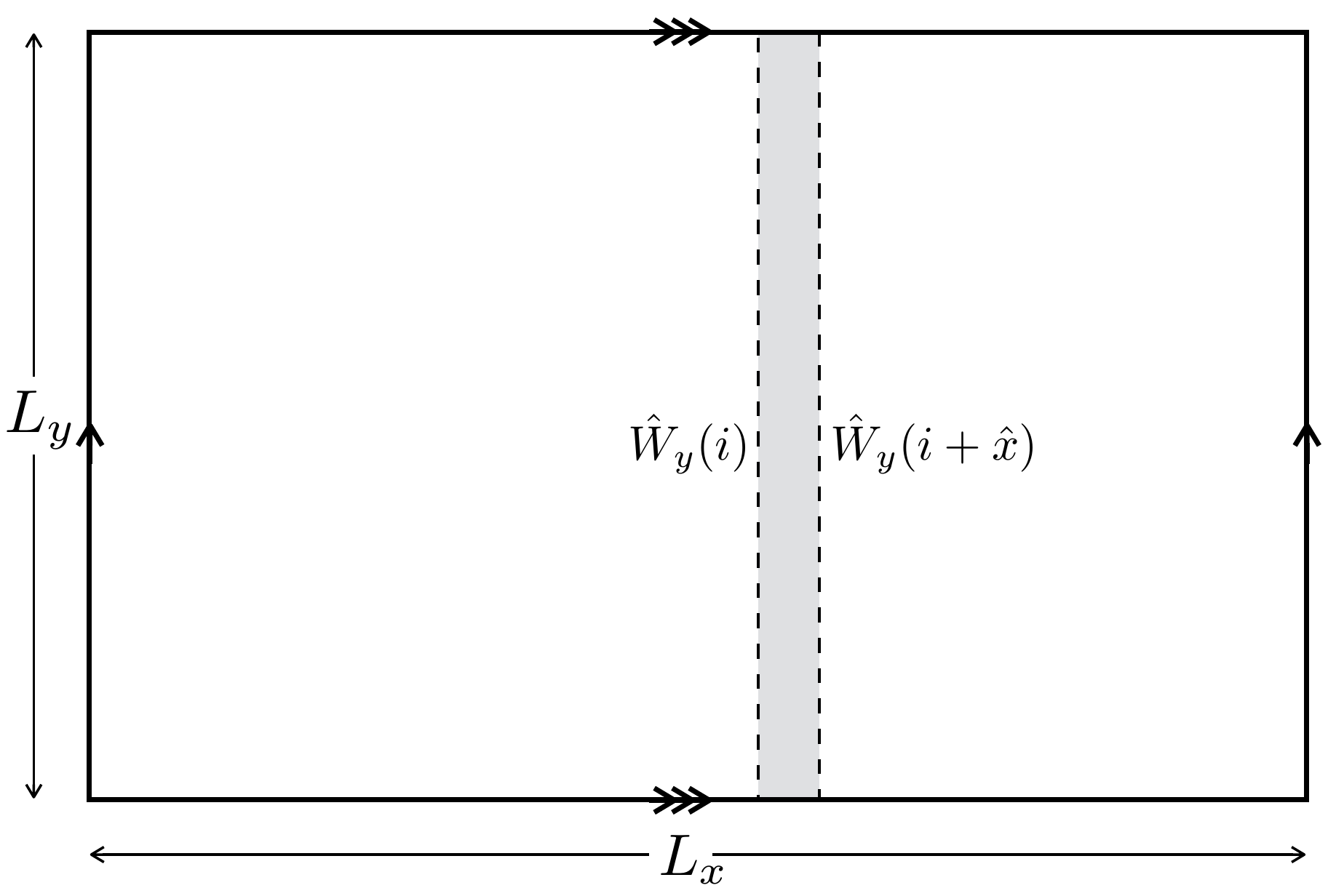}
\end{center}
\caption{Square lattice on a torus. The Wilson loop $\hat{W}_y$ is translated by one lattice spacing in the $\hat{x}$ direction.}
\label{fig:strip}
\end{figure}

The TQFT can also implement the second crucial property of the vison described above, and illustrated in Fig.~\ref{fig:visonberry}.
As in Section~\ref{sec:fl}, we place the $\mathbb{Z}_2$ spin liquid on a square lattice of size $L_x \times L_y$ with toroidal
boundary conditions. Now consider the impact of translation by one lattice spacing, $\hat{T}_x$, on the Wilson loop operator
$\hat{W}_y$, as shown in Fig.~\ref{fig:strip}. 
The motion of the Wilson loop encloses $L_y$ lattice sites, and so this operation is equivalent to a vison having encircled $L_y$
sites. From Section~\ref{sec:sl}, we conclude that such a process yields a Berry phase of $2 \pi S L_y$. The net result is the following
non-trivial operator relation
\beq
\hat{T}_x \hat{W}_y = e^{2 \pi i S L_y} \hat{W}_y \hat{T}_x,
\label{TW}
\eeq
and a second relation with $x \leftrightarrow y$. 
Note that for $L_y$ odd, $\hat{T}_x$ and $\hat{W}_y$ anti-commute (commute) for odd (even) $\mathbb{Z}_2$ spin liquids.
These relationships are not part of the TQFT structure per se, but instead show how
global symmetries of the underlying quantum system are realized in a non-trivial manner by the TQFT operators. In other words,
they describe the `symmetry enriched topological' structure, or the 
`symmetry fractionalization' by gapped excitations, \cite{EH13,WenSPT,YQLF15} of the $\mathbb{Z}_2$ spin liquid.

\subsection{Momentum balance}
\label{sec:mombal}

The general results in (\ref{deltaP}) and (\ref{deltaP2}), describing flux insertion through the cycle of torus,
apply to any lattice quantum system with a global U(1) symmetry, and so should
also apply to the $\mathbb{Z}_2$ spin liquid. We will now show, using (\ref{TW}), that (\ref{deltaP}) and (\ref{deltaP2})  are indeed satisfied.

\begin{figure}
\begin{center}
\includegraphics[height=3.5cm]{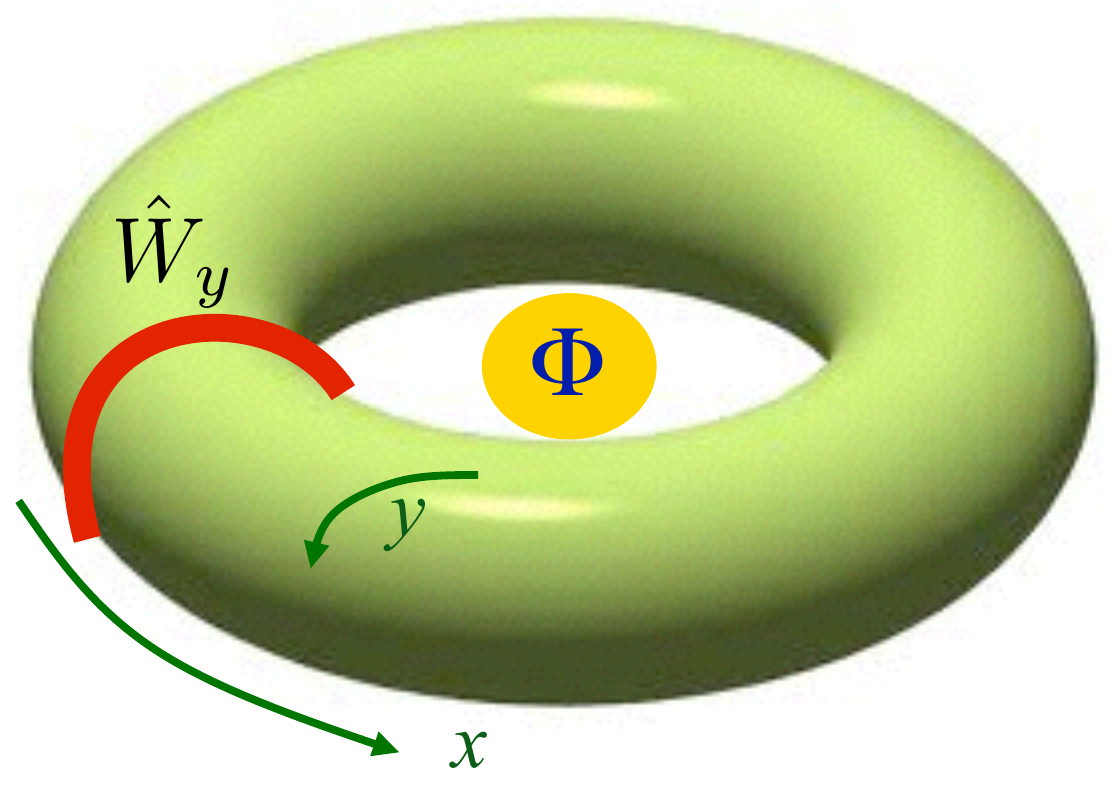}
\end{center}
\caption{As in Fig.~\ref{fig:torus}. For a $\mathbb{Z}_2$ spin liquid, the flux insertion is equivalent to an operator
acting on the red line: this is the branch-cut operator
acting on the RVB state, or equivalently, the operator $\hat{W}_y$ of the TQFT.}
\label{fig:torus2}
\end{figure}

As in Section~\ref{sec:fl}, we insert a flux, $\Phi$, which couples only to the up spin electrons, which requires choosing $A_\mu^s = 2 A_\mu^e \equiv A_\mu$. We work in real time, and thread a flux along the $x$-cycle of the torus. So we have
\beq
A_x = \frac{\Phi (t)}{L_x}
\label{fls2}
\eeq
where $\Phi (t)$ is a function which increases slowly from 0 to $2 \pi$.  In (\ref{SCS}), the $A_x$ gauge field
couples only to $a_y$, and we parameterize
\beq
a_y = \frac{\theta_y}{L_y}.
\label{fls3}
\eeq
Then, from (\ref{SCS}), the time evolution operator of the flux-threading operation can be written as
\beq
\hat{U} = \exp \left( \frac{i}{2 \pi} \int dt \, \hat{\theta}_y \frac{d \Phi}{dt} \right) = e^{i \hat{\theta}_y} \equiv \hat{W}_y
\label{fls4}
\eeq
So the time evolution operator is simply the Wilson loop operator $\hat{W}_y$.
If the state of the system before the flux-threading was $\left| G \right\rangle$, then the state after the
flux threading will be $ \hat{W}_y\left| G \right\rangle$. This is illustrated in Fig.~\ref{fig:torus2}.

Now we can easily determine the difference in momenta of the states $\left| G \right\rangle$ and $ \hat{W}_y\left| G \right\rangle$.
From (\ref{TW}) we obtain
\beq
\Delta P_x = 2 \pi S L_y \, (\mbox{mod} \, 2 \pi) = \frac{2 \pi}{L_x} \left( S L_x L_y \right) \, (\mbox{mod} \, 2 \pi). \label{fls5}
\eeq
In the second form above, we see that (\ref{fls5}) is consistent with (\ref{deltaP2}) for $N_\uparrow =S L_x L_y$. This is indeed the 
correct total number of up spin electrons in a spin $S$ antiferromagnet.

\section{Fermi liquid-like metals with topological order for the pseudogap state}
\label{sec:fls}

A simple picture of the fractionalized Fermi liquid (FL* metal) \cite{TSSSMV03,TSMVSS04,APAV04,CMS05,Bonderson16,Tsvelik16} 
is that it is a combination of the systems described in Sections~\ref{sec:fl}
and~\ref{sec:sl}. The low energy excitations of such a state on a torus are given by the 
action
\beq
\mathcal{S}_{FL^*} = \mathcal{S}_{FL} + \mathcal{S}_{CS}
\eeq
which is the
direct sum of the action for 
fermionic quasiparticles in (\ref{SFL}), and of the action for the TQFT in (\ref{SCS}). Consequently the momentum balance also
involves the direct sum of the quasiparticle contribution in (\ref{Vfs}) and the TQFT contribution in (\ref{fls5}) for $S=1/2$; 
and these should add up to the total number of up spin electrons $N_\uparrow = N/2$ in (\ref{deltaP2}). 
So the presence of an odd $\mathbb{Z}_2$ spin liquid leads to a modified
constraint on the volume of the Fermi surface enclosed by the quasiparticles. For the cuprate case, with a total density of $1+p$
holes, we have in the FL* metal a modification from (\ref{LT}) to 
\beq
\frac{V_{FS}}{2 \pi^2} = p \, (\mbox{mod} \, 2).
\label{LTs}
\eeq

The simplest realizations of FL* are in 2-band Kondo-Heisenberg lattice models \cite{TSSSMV03,TSMVSS04,CMS05,Tsvelik16}.
Then the origin of the direct sum picture described above can be understood in a simplified picture:
the local moments with Heisenberg exchange interactions can form the spin liquid, while the conduction electrons form the `small'
Fermi surface. This simple picture assumes the Kondo exchange between the local and itinerant electrons can be neglected, 
but it can be important for determining whether FL* is realized in a specific model \cite{Tsvelik16}.

However, for the cuprates we need a realization of FL* in a 1-band model, as only a single band of electronic excitations is observed.
Such a realization has appeared in a series of works \cite{CS93,SS94,XGWPAL96,XGW06,RKK07,RKK08,RKK08b,YQSS10,Mei12,MPSS12,Ferraz13,Punk15}. Here we briefly describe the simplified
model of Ref.~\cite{Punk15}, which extends the RVB picture to include mobile fermionic carriers which have the same quantum numbers
as the electron. 
\begin{figure}
\begin{center}
\includegraphics[height=6.5cm]{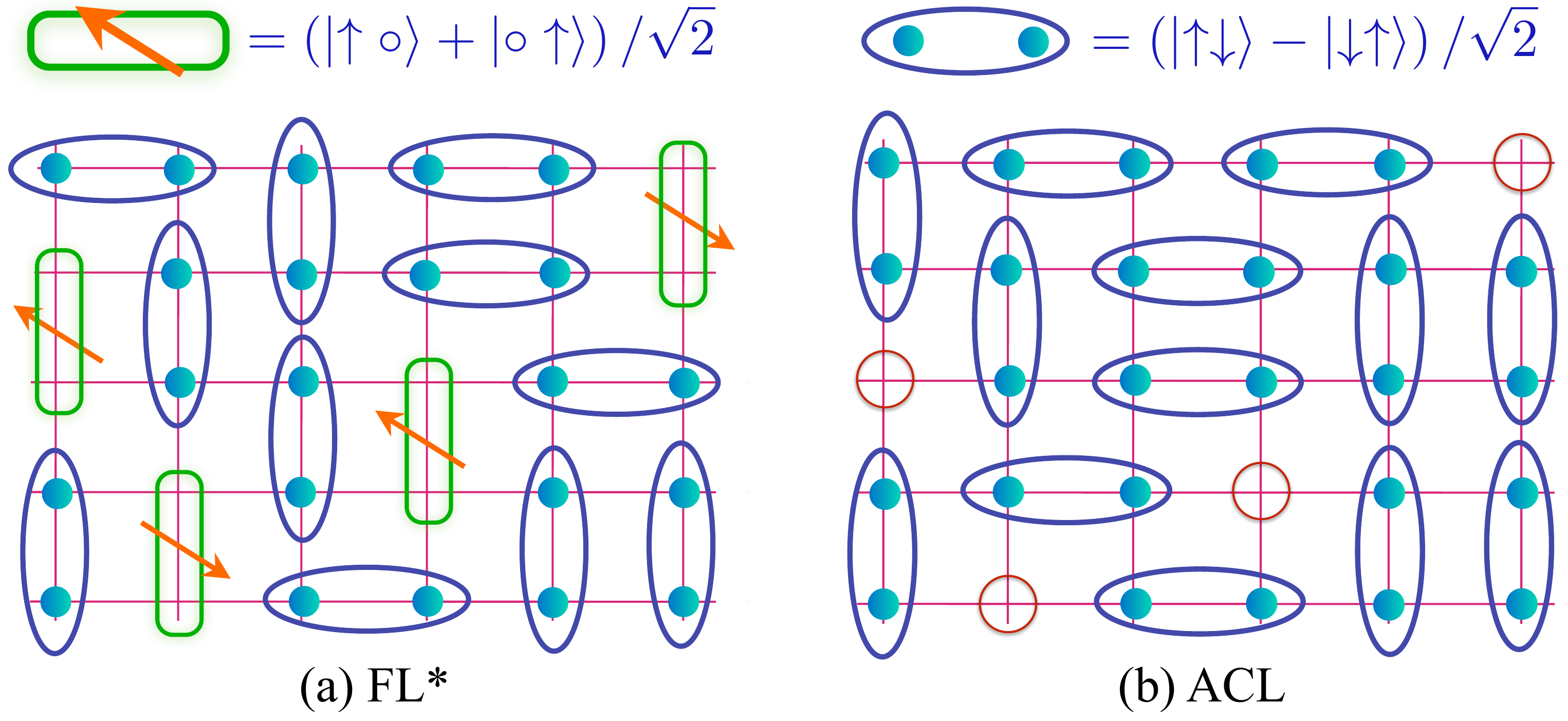}
\end{center}
\caption{(a) A component of a resonating bond wavefunction for FL* in a single-band model on the square lattice \cite{Punk15}. 
The density of the green
bonds is $p$, and these are fermions which form Fermi surface of volume (\ref{LTs}) with electron-like quasiparticles.
(b) A component of a wavefunction for an ACL. The vacancies are the `holons', or more generally, the `chargons'; they are 
assumed to be fermions which form a Fermi liquid-like state with a Fermi surface of 
spinless quasiparticles of charge $e$.}
\label{fig:fls}
\end{figure}
As shown in Fig.~\ref{fig:fls}a, we construct a trial wavefunction as a superposition of valence bond coverings of the 
square lattice with two distinct categories of pairs of sites: ({\em i\/}) the blue bonds in Fig.~\ref{fig:fls}a, which represent a pair of electrons in 
a singlet bond, and ({\em ii\/}) the green bonds in Fig.~\ref{fig:fls}a, which represent a single electron in a bonding orbital between the sites.
The density of green bonds is $p$, and relative to the RVB background of blue bonds, each green bond is a fermion which
carries charge $+e$ and spin $S=1/2$ {\em i.e.\/} the same quantum numbers as a hole in a band insulator. These mobile green fermions can
then form a `small' Fermi surface of volume given by (\ref{LTs}). The background of resonating blue and green bonds still preserves the topological order
of the spin liquid, and forms a sector described by the TQFT of a $\mathbb{Z}_2$ spin liquid \cite{PCAS16}.

We also show in Fig.~\ref{fig:fls}b a related state called the `holon metal' \cite{XGW89,PAL89}, 
or more generally an `algebraic charge liquid' (ACL) \cite{RKK08}.
In this case, in addition to the blue singlet bonds, we have a density, $p$, of spinless, fermionic vacancies (the `holons', or more generally 
the `chargons') each carrying charge $+e$.
Now the chargons can form a Fermi liquid-like state with a 
small Fermi surface of size $p$, but the quasiparticles at the Fermi surface will not be electron-like, as 
they carry only charge but no spin. Note that although the number of quasiparticle states inside the Fermi surface is the `small'
value $p$, determination of the Fermi wavevector requires accounting for the spin or other quantum numbers carried by the chargons:
for the model in Section~\ref{sec:sm}, the chargons also carry a pseudospin index ($s=\pm$) 
which has the same degeneracy as electronic spin.

The momentum balance argument for an ACL works just like for FL*. The chargons carry charge but no spin, and so they
couple to the electromagnetic gauge field $A_\mu^e$. As we saw in Section~\ref{sec:mombal}, flux insertion coupling
only to spin-up particles is carried out using $A_\mu^s = 2 A_\mu^e \equiv A_\mu$, and the net result is that the chargon
contribute just as naively expected: as spinless fermions of charge $+e$, making up a metal with a total charge density
of $p e$ mobile carriers. 
In general, both chargon and electron-like Fermi surfaces can be present, and their sizes should sum to $p$ \cite{RKK08}.

Turning to the phase diagram of the cuprates in Fig.~\ref{fig:phasediag}b, we now summarize the evidence that a
FL* (or an ACL) model describes the PG regime.
\begin{itemize}
\item Model computations \cite{YQSS10,Punk15} of the Fermi surface configuration for FL* yield hole pockets centered near, but not 
exactly at, $(\pi/2,\pi/2)$. The electron spectral weight is highly anisotropic around the Fermi surface and this can possibly
explain the photoemission observation of arc-like regions of significant spectral weight \cite{YangJohnson11}. Similar spectral weights
can also be obtained in models of the ACL \cite{RKK07}, provided the spinon gap is smaller than the temperature.
\item A $T$-independent positive Hall coefficient $R_H$ corresponding to carrier density $p$ in the higher temperature pseudogap \cite{Ando04}. This is the expected Hall co-efficient of the hole pockets in the FL* or ACL phase.
\item The frequency and temperature dependence of the optical conductivity has a Fermi liquid form $\sim 1/(-i \omega + 1/\tau)$ with $1/\tau \sim \omega^2 + T^2$ \cite{Marel13}. This Fermi liquid form is present although the overall prefactor corresponds
to a carrier density $p$.
\item Magnetoresistance measurements obey Kohler's rule \cite{MG14} with $\rho_{xx} \sim \tau^{-1} \left( 1 + a (H \tau )^2 \right)$, again as expected by Fermi pockets of long-lived charge-carrying quasiparticles.
\item Density wave modulations have long been observed in STM experiments \cite{Kohsaka07} 
in the region marked DW in Fig.~\ref{fig:phasediag}b. Following theoretical proposals \cite{MMSS10b,SSRLP13}, a number of experiments \cite{Fujita14,Comin14,Forgan15,MHH15a,MHH15b} have identified the pattern of modulations as a $d$-form factor density wave. Computations of density wave instabilities of the FL* metal lead naturally to a $d$-form factor
density wave, with a wavevector similar to that observed in experiments \cite{DCSS14b,NH16} (very similar results
would be obtained in a computation starting from an ACL metal because the density wave instabilities are not sensitive
to the spin of the quasiparticles).
In contrast, computation of density wave
instabilities of the large Fermi surface FL metal lead to density wave order along a `diagonal' wavevector not 
observed in experiments \cite{SSRLP13,TS15,BS15}.
\item Finally, very interesting recent measurements by Badoux {\em et al.} \cite{LTCP15} 
of the Hall co-efficient at high fields and low $T$ for 
$p \approx 0.16$ in YBCO clearly show the absence of DW order, unlike those at lower $p$. Furthermore unlike the DW
region, the Hall co-efficient remains positive and corresponds to a density of $p$ carriers. Only at higher $p \approx 0.19$
does the FL Hall co-efficient of $1+p$ appear:
in Fig.~\ref{fig:phasediag}b, this corresponds
to the $T^\ast$ boundary extending past the DW region at low $T$. A possible explanation is that the FL* or ACL phase is present at $p=0.16$.
\end{itemize}

\section{Fluctuating antiferromagnetism and the strange metal}
\label{sec:sm}

The strange metal (SM) region of Fig.~\ref{fig:phasediag}b exhibits strong deviations in the temperature and frequency
dependence of its transport properties from those of a Fermi liquid. Its location in the temperature-density phase diagram suggests that the
SM is linked to the quantum criticality of a zero temperature critical point (or phase) near $p=0.19$. 
We interpret the experiments as placing a number of constraints on a possible theory:
\begin{itemize}
\item The quantum transition is primarily ``topological''. The main change is in the size of the Fermi surface from small (obeying (\ref{LTs})) to 
large (obeying (\ref{LT})) with increasing $p$. This is especially clear from the recent Hall effect observations of Badoux {\em et al.\/} \cite{LTCP15}.
\item Symmetry-breaking and Landau order parameters appear to play a secondary role. A conventional order which changes the size of the
Fermi surface must break translational symmetry, and the only such observed order is the charge density wave (DW) order. However, the correlation
length of this order is rather short in zero magnetic field, and in any case it seems to disappear at a doping which is smaller than $p=0.19$;
see Fig.~\ref{fig:phasediag}b.
\item The main symmetry breaking which could be co-incident with the transition at $p=0.19$ is Ising-nematic ordering. But this symmetry cannot change the size of the Fermi surface. Similar comments apply to time-reversal symmetry breaking order parameters that do not break translational symmetry.
\item The small doping side of the critical point exhibits significant spin fluctuations at wavevectors close to but not equal 
to $(\pi,\pi)$, and these become anisotropic
when the Ising-nematic order is present.
\item The Hall effect observations of Badoux {\em et al.\/} \cite{LTCP15} show a smooth evolution of the Hall resistance
between values corresponding to a density of $p$ carriers at $p=0.16$, to that corresponding to a density of $1+p$ 
carriers at $p=0.19$. Such a smooth evolution is very similar to that obtained in a model of the reconstruction of the Fermi surface
by long-range antiferromagnetism \cite{Storey16}. It is possible that magnetic-field-induced 
long-range antiferromagnetism is actually present
in the high field measurements of Badoux {\em et al.\/} \cite{LTCP15}, but fluctuating antiferromagnetism with a large correlation length is a more likely possibility.  We will describe below a model of a ACL/FL* metal 
based upon a theory of fermionic chargons in the presence of local antiferromagnetism without long-range order: 
the evolution of the Hall effect in this model has
little difference from that in the case with long-range antiferromagnetism. We note that theories of a change in Fermi 
surface size involving {\em bosonic\/} chargons \cite{LeeWenRMP,TSMVSS04,Mei12} lead to a 
jump in the Hall co-efficient at the critical point \cite{CMS05} (when the half-filled band of fermionic spinons discontinuously acquires an electromagnetic charge upon the transition from FL* to FL \cite{TSMVSS04}), and this appears to be
incompatible with the data. 
\end{itemize}

It appears we need a gauge theory for a topological transition from a deconfined $\mathbb{Z}_2$-FL* state (or the related
$\mathbb{Z}_2$-ACL state) to a confining FL with a
large Fermi surface involving {\em fermionic\/} chargons.
Significant non-$(\pi,\pi)$ spin correlations should be present in the
deconfined $\mathbb{Z}_2$ state.
Moreover, we would like Ising-nematic order to be present as a spectator of the deconfined $\mathbb{Z}_2$ state,
and disappear at the confining transition to the FL state. 
This situation is the converse of that found in models of `deconfined criticality' \cite{NRSS91,RJSS91,senthil1}, where the spectator order parameter
appears in the confining phase, and not in the deconfined phase wanted here.
Also, the critical point should not be simply given by a theory of the Ising-nematic order, as this cannot account for the charge in the Fermi
surface size.

We now describe a model compatible with these constraints. We will begin with a lattice model of electrons coupled to spin fluctuations. When the spin fluctuations can be neglected, we have a conventional FL state with a large Fermi surface. Conversely, when spin fluctuations are condensed, we
have antiferromagnetic long-range order with small pocket Fermi surfaces. However, our focus will be on possible `deconfined' intermediate
phases where there is no long-range antiferromagnetic order, but the local magnitude of the antiferromagnetic order is nevertheless finite:
the local order determines the magnitude of the pseudogap and leads to small pocket Fermi surfaces even without long-range order.
We will argue that the concept of `local antiferromagnetic order' can be made precise by identifying it with the Higgs phase of an emergent 
gauge theory. The Higgs phase will realize the small Fermi surface 
$\mathbb{Z}_2$ FL* or ACL phases discussed above as models of the PG metal.

The model of electrons coupled to spin fluctuations has the Lagrangian
\beq
\mathcal{L} = \mathcal{L}_c +  \mathcal{L}_{c\Phi} + \mathcal{L}_\Phi.
\label{Lspinfermion}
\eeq
The first term describes the fermions, $c_{i \alpha}$, hopping on the sites of a square lattice.
\beq
\mathcal{L}_c = \sum_i c_{i\alpha}^\dagger \left[\left(\frac{\partial}{\partial \tau}-\mu\right)\delta_{ij}-t_{ij}\right]c_{j\alpha}.
\label{Lc}
\eeq
We describe the interactions between the fermions via their coupling to spin fluctuations at 
the 
wavevectors ${\bm K}_x$ and ${\bm K}_y$ which are close to but not equal to $(\pi,\pi)$, and are related by 90$^\circ$ rotation.
Along ${\bm K}_x$ this is characterized by a complex vector in spin space $\Phi_{x\ell}$, and similarly for
${\bm K}_y$ so that
\beq 
\left\langle c_{i\alpha}^\dagger \sigma^\ell_{\alpha\beta} c_{i\beta} \right\rangle
\sim \Phi_{ix\ell} \, e^{ i {\bm K}_x \cdot{\bm r}_i } + \mbox{c.c.} + \Phi_{iy\ell} \, e^{ i {\bm K}_y \cdot{\bm r}_i } + \mbox{c.c.}.
\label{norder}
\eeq
Then the Lagrangian coupling the electrons $c_{i \alpha}$ to the spin fluctuations is given by
\beq
{\mathcal L}_{c\Phi} = - \lambda\sum_i \left[\Phi_{ix\ell} \, e^{ i {\bm K}_x \cdot{\bm r}_i } + \mbox{c.c.}  +  \Phi_{iy\ell} \, e^{ i {\bm K}_y \cdot{\bm r}_i } + \mbox{c.c.} \right]
\, c_{i\alpha}^\dagger {\sigma}^\ell_{\alpha\beta} c_{i\beta} .
\label{model}
\eeq
The coupling $\lambda$ is expected to be large, and our discussion below will implicitly assume so.
Finally, we have the Lagrangian describing the spin fluctuations 
\beq
\mathcal{L}_\Phi =   \left[ |\partial_\tau \Phi_{x\ell} |^2 + v^2 |\nabla \Phi_{x\ell}|^2 + |\partial_\tau \Phi_{y\ell} |^2 + v^2 |\nabla \Phi_{y\ell}|^2 
 +  r \left( |\Phi_{x\ell}|^2 + |\Phi_{y \ell}|^2 \right) + \ldots \right],
\eeq
where $v$ is a spin-wave velocity.

The theory $\mathcal{L}$ is often referred to as a `spin-fermion' model \cite{ACS03}, and it 
provides the theory for the direct onset of antiferromagnetism in a Fermi liquid.
There has been a great deal of work on this topic, starting with the work of Hertz \cite{Hertz76,Millis93,abanov00,MMSS10b}. 
Recent  sign-problem-free quantum Monte Carlo simulations of spin-fermion models \cite{BMS12,SGTB15,LWYL15,LWYL16,DSSV15}
have yielded phase diagrams with remarkable similarities to those of the pnictides and the electron-doped cuprates.
The spin-fermion problem can also be applied at half-filling ($p=0$) with $(\pi, \pi)$ antiferromagnetic correlations,
and then the background half-filled density of $c$ fermions 
yields \cite{FSX11} the correct Berry phases of the `hedgehog' defects in the 
N\'eel order parameter \cite{Haldane88,NRSS89,NRSS90}. The latter Berry phases are characteristic of the insulating Heisenberg
antiferromagnet
at $p=0$, and so  a judicious treatment of the spin-fermion model at large $\lambda$ can also describe Mott-Hubbard physics.

Here, we want to extend the conventional theoretical treatments of the spin-fermion model to reach more exotic states with
Mott-Hubbard physics at non-zero $p$. The formalism we present below can yield insulators at $p=0$
both with and without AF order, with the latter being topological phases with emergent gauge fields. Moreover, the topological order
will also extend to metallic phases at non-zero $p$, with gauge-charged Higgs fields describing local antiferromagnetism in the presence
of a pseudogap and small pocket Fermi surfaces, 
but without long-range antiferromagnetic order. 

The key step in this process \cite{SS09,DCSS15b,DCSS15} is to transform the electrons to a rotating reference frame
along the local magnetic order, using a SU(2) rotation $R_i$ and (spinless-)fermions 
$\psi_{i,s}$ with $s= \pm$,
\beq
\left( \begin{array}{c} c_{i\uparrow} \\ c_{i\downarrow} \end{array} \right) = R_i \left( \begin{array}{c} \psi_{i,+} \\ \psi_{i,-} \end{array} \right),
\label{R}
\eeq
where $R_i^\dagger R_i = R_i R_i^\dagger = 1$. Note that this representation immediately introduces a SU(2) gauge invariance (distinct from the global SU(2) spin rotation)
\beq
\left( \begin{array}{c} \psi_{i,+} \\ \psi_{i,-} \end{array} \right)\rightarrow U_i (\tau) \left( \begin{array}{c} \psi_{i,+} \\ \psi_{i,-} \end{array} \right) \quad, \quad R_i \rightarrow R_i U_i^\dagger (\tau), \label{gauge}
\eeq
under which the original electronic operators remain invariant, $c_{i\alpha}\rightarrow c_{i\alpha}$; here $U_i (\tau) $ is a  SU(2) gauge-transformation acting on the $s=\pm$ index. So the $\psi_s$ fermions are SU(2) gauge fundamentals, carrying 
the physical electromagnetic global U(1) charge, but not the SU(2) spin of the electron: they are the fermionic  
``chargons'' of this theory, and  the density of the $\psi_s$ is the same as that of the electrons.
The bosonic $R$ fields also carry the global SU(2) spin (corresponding to left multiplication of $R$) but are electrically neutral:
they are the bosonic ``spinons'', and are related \cite{NRSS89,SSNR91,SS09} to the Schwinger bosons in (\ref{Hb}).
Later, we will also find it convenient to use the parameterization
\beq
R = \left(
\begin{array}{cc} z_\uparrow & - z_{\downarrow}^\ast \\
z_{\downarrow} & z_\uparrow^\ast \end{array} \right) \label{Rz}
\eeq
with $|z_\uparrow |^2 + |z_\downarrow |^2 = 1$.

A summary of the charges carried by the fields in the resulting SU(2) gauge theory, $\mathcal{L}_g$, is in Table~\ref{tab:charge}.
\begin{table}
\begin{center}
\begin{tabular}{|c|c|c|c|c|c|}
\hline
Field & Symbol & Statistics & SU(2)$_{\rm gauge}$ & SU(2)$_{\rm spin}$ & U(1)$_{\rm e.m. charge}$ \\
\hline 
Electron & $c$ & fermion & ${\bm 1}$ & ${\bm 2}$ & -1\\
AF order & $\Phi$ & boson & ${\bm 1}$ & ${\bm 3}$ & 0 \\
Chargon & $\psi$ & fermion & ${\bm 2}$ & ${\bm 1}$ & -1 \\
Spinon & $R$ or $z$ & boson & $\bar{\bm 2}$ & $ {\bm 2}$ & 0 \\
Higgs & $H$ & boson & ${\bm 3}$ & ${\bm 1}$ & 0 \\
\hline
\end{tabular}
\end{center}
\caption{Quantum numbers of the matter fields in $\mathcal{L}$ and $\mathcal{L}_g$. 
The transformations under the SU(2)'s are labelled by the dimension
of the SU(2) representation, while those under the electromagnetic U(1) are labeled by the U(1) charge.
The antiferromagnetic spin correlations are characterized by $\Phi$ in (\ref{norder}).
The Higgs field determines local spin correlations via (\ref{e2}).}
\label{tab:charge}
\end{table}
This rotating reference frame perspective was used in the early work by Shraiman and Siggia on lightly-doped
antiferromagnets \cite{SS88,SS89}, although their attention was restricted to phases with antiferromagnetic order. The importance of the 
gauge structure
in phases without antiferromagnetic order was clarified in Ref.~\cite{SS09}.

Given the SU(2) gauge invariance associated with (\ref{R}), when we express $\mathcal{L}$ in terms of $\psi$ we naturally obtain a SU(2)
gauge theory with an emergent gauge field $A_\mu^a =(A_\tau^a ,{\bm A}^a)$, with $a=1,2,3$. 
We write the Lagrangian of the resulting gauge theory as \cite{SS09,DCSS15b,DCSS15}
\beq
\mathcal{L}_{g} = \mathcal{L}_\psi + \mathcal{L}_Y + \mathcal{L}_R + \mathcal{L}_H  .
\label{LQCP}
\eeq
The first term for the $\psi$ fermions descends directly from the $\mathcal{L}_c$ for the electrons
\beq
\mathcal{L}_\psi = \sum_i \psi_{i,s}^\dagger \left[\left(\frac{\partial}{\partial \tau}-\mu \right)\delta_{ss'} + iA_\tau^a\sigma^a_{ss'} \right]\psi_{i,s'} +  \sum_{i,j}t_{ij}\psi^\dagger_{i,s}\bigg[e^{i\sigma^a {\bm A}^a\cdot({\bm r}_i-{\bm r}_j)}\bigg]_{ss'}\psi_{j,s'},
\eeq
and uses the same hopping terms for $\psi$ as those for $c$, along with a minimal coupling to the SU(2) gauge field.
Inserting (\ref{R}) into $\mathcal{L}_{cn}$, we find that the resulting expression involves 2 complex Higgs fields, $H_x^a$ and $H_y^a$, which
are SU(2) adjoints; these are defined by
\beq
H^a_x \equiv \frac{1}{2} \Phi_{x\ell} \, \mbox{Tr}[\sigma^\ell R\sigma^a R^\dagger],
\label{hnc}
\eeq
and similarly for $H^a_y$.  Let us also note the inverse of (\ref{hnc})
\beq
\Phi_{x\ell} = \frac{1}{2} H^a_x \, \mbox{Tr}[\sigma^\ell R\sigma^a R^\dagger],
\label{e2}
\eeq
and similarly for $H^a_y$, expressing the antiferromagnetic spin order in terms of the Higgs fields and $R$.
Then $\mathcal{L}_{c\Phi}$ maps to the form of a `Yukawa' coupling equal to $\lambda$, 
\beq
\mathcal{L}_Y = -  \lambda \sum_i  \left( H_{ix}^{a}  e^{i {\bm K}_x \cdot {\bm r}_i } + H_{ix}^{a \ast}  e^{-i {\bm K}_x \cdot {\bm r}_i } 
+ H_{iy}^{a}  e^{i {\bm K}_y \cdot {\bm r}_i } + H_{iy}^{a \ast}  e^{-i {\bm K}_y \cdot {\bm r}_i } \right) \, \psi_{i,s}^\dagger \sigma^a_{ss'} \psi_{i,s'}. \label{yukawac}
\eeq
We note again that our discussion below will implicitly assume large $\lambda$.
The remaining terms in the Lagrangian involving the bosonic Higgs field, $H$, the bosonic spinons $R$, and the gauge field $A_\mu^a$ follow
from gauge invariance and global symmetries, and are similar to those found in theories of particle physics. In particular the spinon Lagrangian is
\beq
\mathcal{L}_R =  \frac{1}{2g}\mbox{Tr}\bigg[(\partial_\tau R - iA_\tau^aR\sigma^a)(\partial_\tau R^\dagger + iA_\tau^a\sigma^aR^\dagger) + v^2(\nabla R - i{\bm A}^aR\sigma^a)(\nabla R^\dagger + i{\bm A}^a\sigma^aR^\dagger) \bigg]. \label{LR}
\eeq
For the Higgs field, we have
\bea
\mathcal{L}_H &=& \left|\partial_\tau H_x^a + 2i\epsilon_{abc}A^b_\tau H_x^c\right|^2 + \tilde{v}^2 \left|\nabla H_x^a + 2i\epsilon_{abc}{\bm A}^b H_x^c\right|^2 \nn
&+&  \left|\partial_\tau H_y^a + 2i\epsilon_{abc}A^b_\tau H_y^c\right|^2 + \tilde{v}^2 \left|\nabla H_y^a + 2i\epsilon_{abc}{\bm A}^b H_y^c\right|^2
+V(H),
\label{LH}
\eea
with the Higgs potential
\bea
V(H) &=&  h \left( |H_x^a|^2 + |H_y^a|^2 \right) + 
u_1 \left(  [H_x^{a \ast} H_x^a]^2 + [H_y^{a \ast} H_y^a]^2 \right) + u_2 \left( [H_x^{a}]^2 [H_x^{b\ast}]^2 + [H_y^{a}]^2 [H_y^{b\ast}]^2 \right) \nn
&~&~~~~~+ u_3 [H_x^{a \ast} H_x^a][H_y^{b \ast} H_y^b] + u_4 [H_x^{a \ast} H_x^b][H_y^{a \ast} H_y^b]
+ u_5 [H_x^{a \ast} H_x^b][H_y^{b \ast} H_y^a]. \label{VH}
\eea

Despite the apparent complexity of the gauge theory Lagrangian, $\mathcal{L}_g$, described above, it should be noted that its structure
follows largely from the quantum number assignments in Table~\ref{tab:charge}, and the transformations of the fields under
lattice translation. Under translation by a lattice vector ${\bm r}$, $c$, $\psi$, and $R$ transform trivially, while
\beq
H_x^a \rightarrow H_x^a e^{i {\bm K}_x \cdot {\bm r}} \quad , \quad H_y^a \rightarrow H_y^a e^{i {\bm K}_y \cdot {\bm r}}
\eeq
(and similarly for $\Phi_{x\ell}$ and $\Phi_{y\ell}$). The physical interpretations are obtained 
from the mappings
in (\ref{norder}), (\ref{R}), and (\ref{e2}) between the physical observables and the gauge-charged fields in Table~\ref{tab:charge}. 
Also note that, while we can
take the continuum limit for the bosonic fields, the fermionic fields have to be described on a lattice to account for the Fermi surface
structure. 

The main innovation of the above description \cite{SS09,DCSS15b,DCSS15} is the introduction of the Higgs fields
$H^a_x$ and $H^a_y$ as a measure of the local antiferromagnetic order along wavevectors ${\bm K}_x$ and ${\bm K}_y$.
As these Higgs fields only carry SU(2) gauge charges (see Table~\ref{tab:charge}), their condensation does not
break the global SU(2) spin rotation symmetry. However, their magnitude is a gauge invariant observable, and this does
measure the magnitude of the local `pseudogap' created by the Higgs condensate, and changes the dispersion of the fermionic charge
carriers into small pocket Fermi surfaces. 
So the Higgs phase, with no other fields condensed, will realize the PG metal in a theory of local antiferromagnetic correlations.

\subsection{Phase diagram}

We now return to the full model with SU(2) spin rotation symmetry, and discuss the possible phases of the SU(2) gauge theory 
$\mathcal{L}_g$. 

\begin{figure}
\begin{center}
\includegraphics[height=8.5cm]{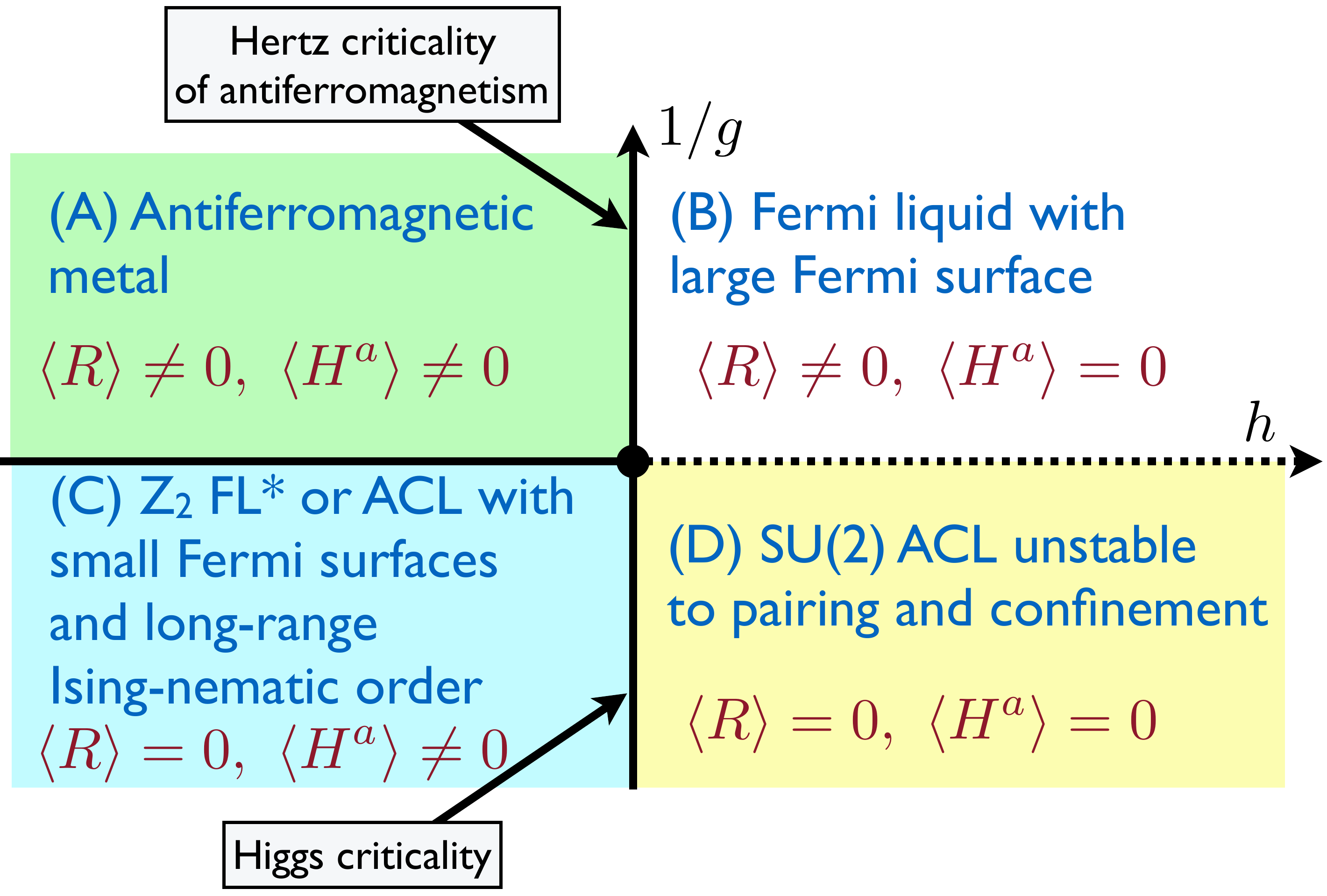}
\end{center}
\caption{Mean-field phase diagram of the SU(2) gauge theory $\mathcal{L}_g$, as a function
of the coupling $g$ in (\ref{LR}) and the `mass' $h$ in (\ref{VH}). 
Phase A has antiferromagnetic order at wavevectors close to, but not equal to, $(\pi, \pi)$.
Phase C is our candidate for the PG metal, and
the `Higgs criticality', between phases C and D, is our candidate for the description of the strange metal. The boundary
between phases B and D does not remain a phase transition \cite{caveat} after confinement of the SU(2) gauge theory 
has been accounted for: the boson $R$ carries fundamental SU(2) gauge charge, and its Higgs (B) 
and confinement (D) phases 
are smoothly connected \cite{FS79}. 
This phase diagram shows how the conventional physics of Hertz criticality, applicable to the pnictides, evolves 
naturally
to the `topological' physics of Higgs criticality, applicable to the hole-doped cuprates.
The Fermi surface reconstruction across the B-A boundary is due to the
antiferromagnetic order, and it involves changes in the band structure of electron-like quasiparticles, $c$. 
A nearly identical Fermi surface reconstruction, with similar transport properties, 
takes place across the D-C boundary, except that it involves
spinless chargons, $\psi$.}
\label{fig:higgs}
\end{figure}

It is useful to proceed by sketching the mean field phase diagram in terms of possible condensates of the bosons $R$ and $H$,
and follow it by an analysis of the role of gauge fluctuations. Such a phase diagram is sketched in Fig.~\ref{fig:higgs} as a function
of the coupling $g$ in (\ref{LR}) and the `mass' $h$ in (\ref{VH}).

\noindent
\underline{Phases A and B:}
The phases in which the spinon, $R$, is condensed are the familiar Fermi liquid phases. This is evident from (\ref{R}), which implies that
with $R$ condensed $c \sim \psi$; also from (\ref{hnc}) the Higgs fields $H$ is related by a global rotation to the antiferromagnetic
order parameter $\Phi$. Consequently, the phase B in Fig.~\ref{fig:higgs} is the conventional Fermi liquid with a large Fermi surface of size $1+p$.
The condensation of $R \sim \Phi$ leads to the onset of antiferromagnetic order in phase A via a Hertz type critical 
point \cite{Hertz76,Millis93,abanov00,MMSS10b}; this condensation will reconstruct the Fermi surface to yield a Fermi liquid with
`small' Fermi surfaces.

We therefore turn our attention to the possibly exotic phases C and D in Fig.~\ref{fig:higgs}.

\noindent
\underline{Phase D:}
There is no Higgs condensate in phase D, and so all the SU(2) gauge fields are active. The gauge-charged 
matter sector includes a large Fermi surface of $\psi_\pm$ fermions which transform as a SU(2) doublet.
The attractive SU(2) gauge force is expected to pair these fermions, leading to a superconducting state \cite{Son99,MMSS14}. 
The resulting gapping
of the fermionic excitations will unscreen the SU(2) gauge force, which will confine all gauge-charged excitations.
Ultimately, we therefore expect phase D to be a superconductor without topological order or fractionalized excitations,
and a conventional Fermi liquid could appear in a magnetic field or at higher temperatures. Also, as indicated in Fig.~\ref{fig:higgs}, we expect phase D to be smoothly connected to the Fermi liquid phase B, as the latter is also unstable to pairing induced by the spin fluctuations \cite{caveat}.

\noindent
\underline{Phase C:}
Finally, we turn our attention to phase C. Here we have a $H$ condensate, and this will break the SU(2) gauge invariance
down to a smaller gauge group. But, because $R$ is not condensed, by (\ref{norder}), global spin rotation invariance is nevertheless preserved.
The case of particular interest to us here is a residual gauge invariance of ${\rm SU}(2)/{\rm SO}(3) \cong \mathbb{Z}_2$.
This will be the situation as long as the Higgs potential $V(H)$ in (\ref{VH}) is such that all four of the real 3-vectors 
$ \langle \mbox{Re} (H_x^a) \rangle$, $ \langle \mbox{Im} (H_x^a) \rangle$, $\langle \mbox{Re} (H_y^a) \rangle$, $ \langle \mbox{Im} (H_y^a) \rangle$ are not parallel to each other, so that the Higgs condensate transforms under SO(3) global rotations.
(For the case with all four vectors parallel, there is a residual U(1) gauge invariance
associated with rotations about the common direction \cite{SS09,DCSS15b}.) A simple case with residual $\mathbb{Z}_2$ gauge
invariance is
\beq
H^a_x \sim (1,i, 0) \quad, \quad H^a_y = (0,0,0), \label{spiralH}
\eeq
or its global SO(3) rotations.
We then obtain an effective $\mathbb{Z}_2$ gauge theory, with the same structure as the TQFT of Section~\ref{sec:sl}.
In particular, the $\pi_1 ({\rm SO}(3)) = \mathbb{Z}_2$ vortices in the 
Higgs field $H$ correspond to vison excitations \cite{NRSS91,SSNR91,SSRMP}, which are gapped in phase C.

For the Higgs condensate in (\ref{spiralH}), writing the spinons $R$ as in (\ref{Rz}),
 (\ref{e2}) becomes
\beq
\Phi_{x\ell} = - \varepsilon_{\alpha\gamma} z_\gamma \sigma^{\ell}_{\alpha\beta} z_\beta, \label{Phiz}
\eeq
where $\varepsilon$ is the unit anti-symmetric tensor (the $z_\alpha$ spinons are connected \cite{NRSS89,SSNR91,SS09} to the Schwinger bosons in (\ref{Hb})). In terms of the real and imaginary components of $\Phi_{x \ell} = n_{1 \ell} + i n_{2  \ell}$, (\ref{Phiz}) yields a pair of orthonormal vectors $n_{1 \ell}$ and $n_{2 \ell}$, describing the SO(3) antiferromagnetic
order parameter. So by (\ref{norder}), (\ref{spiralH})
represents spiral spin correlations along the wavevector ${\bm K}_x$, with no corresponding correlations
along ${\bm K}_y$. Such a state has long-range Ising-nematic order, as correlations of spin-rotation invariant observables
will be different along the $x$ and $y$ lattice directions. 

Our description of phase C so far leads to a $\mathbb{Z}_2$-ACL state, described earlier in simple terms in Section~\ref{sec:fls}.
From the Yukawa coupling in (\ref{yukawac}),
the Higgs condensate reconstructs the Fermi surface of the $\psi$ fermions into a filled band along with 
small pockets: this reconstruction has an identical structure to that of the $c$ fermions across the B-A phase boundary,
and so charge transport across the D-C transition should be similar to that across the B-A transition \cite{DCSS15b}.
The filled band in phase C corresponds to a density of a unit $\mathbb{Z}_2$ gauge charge on every site, 
and so this phase is described by an odd $\mathbb{Z}_2$ gauge theory.
The quasiparticles around the
Fermi surfaces of the small pockets are the $\psi$ chargons (see Table~\ref{tab:charge}), 
and we obtain the ACL state
represented earlier in Fig.~\ref{fig:fls}b.
To obtain a $\mathbb{Z}_2$-FL* state (see Fig.~\ref{fig:fls}a), 
we need the reconstructed $\psi$ quasiparticles to form bound states with the $R$ spinons,
and for the resulting bound state of electron-like quasiparticles to form a Fermi surface: the 
hopping $t_{ij}$ in (\ref{Lc}) is an attractive interaction between the chargons and spinons which can lead to such
bound states.
In general, both $\psi$ and $\psi$-$R$ Fermi surfaces
will be present \cite{RKK08}, and their combined size is restricted by the Luttinger constraint \cite{PSB05,CPR05}.
Computations of models of this bound-state formation have been presented elsewhere \cite{RKK07,RKK08,YQSS10,MPSS12}.
Transport measurements on the PG metal do not distinguish between $\psi$ and $\psi$-$R$ quasiparticles, as they are only sensitive
to the charge carried by the fermionic quasiparticles. However, photoemission only sees $\psi$-$R$ quasiparticles, and more detailed
photoemission observations could determine the situation in the cuprates.

Finally, we comment on the Higgs criticality between phases C and D. Here the theory consists of a critical Higgs field tuned
to the edge of the Higgs phase, by taking the `mass' $h$ in $V(H)$ to its critical value. Because the Higgs condensate is absent,
the $\psi$ fermions form a large Fermi surface, and there is no Ising-nematic order. 
There could also be a spectator small Fermi surface of $\psi$-$R$ quasiparticles,
but this is not expected to be important for the critical theory. The $R$ spinons are gapped, and can also be neglected in the critical theory.
So the final proposed theory for the SM is a large Fermi surface of $\psi$ chargons and a 
critical Higgs field coupled to SU(2) gauge field. Such a theory includes the quantum fluctuations of visons, and their Berry phases,
as it allows for amplitude fluctuations of the Higgs fields, and the lines of zeros in the Higgs field 
correspond to the $\pi_1 ({\rm SO}(3)) = \mathbb{Z}_2$ vortices.
Transport properties of such a theory, and their connection to experiments in the cuprates,
have been discussed recently elsewhere \cite{DCSS15b}. Note that in this scenario, the SU(2) gauge excitations are not deconfined
in either phase C or phase D, and only apparent in the non-Fermi liquid behavior in the finite-temperature quantum critical region \cite{MMSS14};
so this is an example of `deconfined criticality' \cite{senthil1}.

\subsection{Simplified $\mathbb{Z}_2$ and U(1) lattice gauge theories}
\label{sec:Z2}

A notable feature of the phase diagram in Fig.~\ref{fig:higgs} is that none of the ground state phases have deconfined SU(2) 
electric gauge charges, which appear only in a deconfined quantum critical region at non-zero temperature. 
However, deconfined $\mathbb{Z}_2$ electric  
gauge charges are present in phase C. This raises the question of whether it is possible to formulate the theory purely as a $\mathbb{Z}_2$ gauge theory. As was shown in recent work \cite{SBCS16}, it is indeed possible to do so.
The new formulation is defined on the square lattice, and it does not 
yield a direct route to a continuum theory for possible quantum critical points towards confinement. 
Continuum formulations of confinement transitions in
$\mathbb{Z}_2$ gauge theories require 
duality transforms to vison fields via mutual Chern-Simons terms \cite{CXSS09}, but we will not discuss this duality here; it is possible that such an analysis of the criticality will lead back to the deconfined SU(2) gauge theory discussed above.

For simplicity, we consider the case with spiral spin correlations only along the wavevector ${\bm K}_x$; it is not difficult to extend the action below to also include the ${\bm K}_y$ direction. We assume the Higgs field is quenched as in (\ref{spiralH}),
and so write the antiferromagnetic order parameter $\Phi_{x \ell}$ using (\ref{Phiz}). Then the action for the $\mathbb{Z}_2$ lattice gauge theory is \cite{SBCS16}
\beq
\mathcal{L}_{\mathbb{Z}_2} =  \mathcal{L}_c + \mathcal{L}_{cz} + \mathcal{L}_z + \mathcal{L}_\mu , \label{SZ2}
\eeq
where the electron Lagrangian $\mathcal{L}_c$ was specified in (\ref{Lc}), and the coupling between the electrons and the spinons
$z_\alpha$ is obtained by combining (\ref{model}) and (\ref{Phiz})
\beq
\mathcal{L}_{cz} = - \lambda\sum_i \left[ \,  - \varepsilon_{\alpha\gamma} z_{i\gamma} \sigma^{\ell}_{\alpha\beta} z_{i\beta} \, e^{ i {\bm K}_x \cdot{\bm r}_i } + \mbox{c.c.}  \right]
\, c_{i\alpha}^\dagger {\sigma}^\ell_{\alpha\beta} c_{i\beta} .
\label{Lcz}
\eeq
The spinons have the Lagrangian
\beq
\mathcal{L}_z = \frac{1}{g} \left| \partial_\tau z_\alpha \right|^2 - \frac{v^2}{g} \sum_{\langle ij \rangle} \mu^z_{ij} \left( z_{i \alpha}^\ast z_{j \alpha} + \mbox{c.c.}
\right) ,
\label{Lz}
\eeq
where we have introduced an Ising spin, $\mu^z_{ij}=\pm 1$, on the links of the square lattice as a $\mathbb{Z}_2$ gauge field. This gauge field is necessary
because the $z_\alpha$ spinon carries a $\mathbb{Z}_2$ gauge charge.
Finally, we give an independent dynamics to the $\mathbb{Z}_2$ gauge fields, via a standard \cite{KogutRMP} $\mathbb{Z}_2$ gauge theory
Hamiltonian $\mathcal{H}_\mu$, associated with the Lagrangian
$\mathcal{L}_\mu$ in (\ref{SZ2})
\beq
\mathcal{H}_\mu = - K \sum_{\square} \left[ \prod_{\square} \mu^z_{ij} \right]  - h \sum_{\langle ij \rangle} \mu^x_{ij},
\label{SE}
\eeq
where $\mu^x_{ij}$ is a Pauli matrix which anti-commutes with $\mu^z_{ij}$. 
The theory $\mathcal{L}_{\mathbb{Z}_2}$ in (\ref{SZ2}) can be viewed as
a reformulation of the spin-fermion model in (\ref{Lspinfermion}), using additional $\mathbb{Z}_2$ gauge degrees of freedom
that allow for the possibility of fractionalized phases. For small $K$ in (\ref{SE}), we can trace over the $\mathbb{Z}_2$ gauge degrees of
freedom in powers
of $K$, and obtain terms with same structure as in the spin-fermion model in (\ref{Lspinfermion}). 
The unusual feature of the degrees of freedom in $\mathcal{L}_{\mathbb{Z}_2}$, not present in earlier 
treatments \cite{TSMPAF00,LeeWenRMP,FG04,SS09,FSX11,DCSS15b},
is partial fractionalization:
gauge charges are only explicitly present in the spinon sector, while the charged degrees of freedom are 
gauge-invariant electrons.

The main point of Ref.~\onlinecite{SBCS16} is that, 
despite the partial fractionalization in the presentation of $\mathcal{L}_{\mathbb{Z}_2}$, 
the large $K$ and $\lambda$ 
phases of $\mathcal{L}_{\mathbb{Z}_2}$ have the same topological order and fractionalization as those reviewed earlier in the present paper.
At large $K$, $\pi_1 ({\rm SO}(3)) = \mathbb{Z}_2$
vortices in the antiferromagnetic order parameter are suppressed, and this leads to 
phases with $\mathbb{Z}_2$ fractionalization \cite{SSRMP}. At $p=0$, insulating $\mathbb{Z}_2$ spin liquids
like those discussed in Section~\ref{sec:sl} can appear.
In the degrees of freedom in (\ref{SZ2}), the fermionic chargon, $\psi$,
is a bound state of $c$ and $z$ (via $ \psi = R^{-1} c$ from (\ref{R}) and (\ref{Rz})), and its formation can be established in a  large $\lambda$ perturbation theory \cite{SBCS16}. Although the $\mathbb{Z}_2$ gauge sector in
(\ref{SE}) appears to be even, we noted above that the $\mathbb{Z}_2$ fractionalized phase C (in Fig.~\ref{fig:higgs}) has a background filled band of $\psi$ fermions carrying $\mathbb{Z}_2$ electric charges (doping this band leads to small Fermi surfaces), 
and this converts it to $\mathbb{Z}_2$-odd \cite{APAV04}, as was required in Section~\ref{sec:fls} for a small Fermi surface.
So at $p=0$, $\mathcal{L}_{\mathbb{Z}_2}$ describes a Mott insulator with odd $\mathbb{Z}_2$ topological order, similar to those
described in Section~\ref{sec:sl}. At non-zero $p$, at large $\lambda$ and large $K$, 
$\mathcal{L}_{\mathbb{Z}_2}$ exhibits
the fractionalized phase C with all the same characteristics as the SU(2) theory; the conventional phases A and B in Fig.~\ref{fig:higgs} appear at small $K$.
Phase D of the SU(2) gauge theory in Fig.~\ref{fig:higgs} is smoothly connected to phase B, and it does not appear initially as a separate
phase in the $\mathbb{Z}_2$ gauge theory. Finally, the transition from phase C to phase B/D will appear as a confinement
transition in the $\mathbb{Z}_2$ gauge theory upon decreasing $K$, at the same time as the gauge theory changes 
from $\mathbb{Z}_2$-odd to $\mathbb{Z}_2$-even \cite{SBCS16}.

We close this subsection by noting in passing the generalization of $\mathcal{L}_{\mathbb{Z}_2}$ to the case of the U(1) gauge theory of 
collinear antiferromagnetism considered in Refs.~\cite{SS09,DCSS15b}. Now the potential $V(H)$ in (\ref{VH}) is such that the 
Higgs condensates 
are all collinear, and we choose $\langle H_x^a \rangle \sim (0,0,1)$ and $\langle H_y^a \rangle = (0,0,0)$.
The antiferromagnetic order parameter is $\Phi_{x\ell} = z_\alpha^\ast \sigma^{\ell}_{\alpha\beta} z_\beta$,
from (\ref{e2}) and (\ref{Rz}), and this is invariant under the U(1) gauge transformation $z_\alpha \rightarrow e^{i \phi} z_{\alpha}$. The Lagrangian for the U(1) gauge theory, replacing (\ref{SZ2}), is \cite{SBCS16}
\beq
\mathcal{L}_{U(1)} =  \mathcal{L}_c + \mathcal{L}_{cz} + \mathcal{L}_z + \mathcal{L}_A , \label{SZ3}
\eeq
where $\mathcal{L}_c$ remains as in (\ref{Lc}), $\mathcal{L}_{cz}$ in (\ref{Lcz}) is replaced by
\beq
\mathcal{L}_{cz} = - \lambda\sum_i \left[ z_{i \alpha}^\ast \sigma^{\ell}_{\alpha\beta} z_{i\beta} \, e^{ i {\bm K}_x \cdot{\bm r}_i }  + \mbox{c.c.} \right]
\, c_{i\alpha}^\dagger {\sigma}^\ell_{\alpha\beta} c_{i\beta}.
\label{Lcz2}
\eeq
Strictly speaking, such a parameterization applies only at commensurate ${\bm K}_x$, including the case with N\'eel order at ${\bm K}_x = (\pi, \pi)$;
the case with incommensurate collinear antiferromagnetism has an additional `sliding charge mode' \cite{SBCS16}, which we do not treat here.
The spinon Lagrangian $\mathcal{L}_z$ in (\ref{Lz}) is replaced by
\beq
\mathcal{L}_z = \frac{1}{g} \left| \partial_\tau z_\alpha \right|^2 - \frac{v^2}{g} \sum_{\langle ij \rangle} \left( e^{i A_{ij}} z_{i \alpha}^\ast z_{j \alpha} + \mbox{c.c.}
\right),
\label{Lz2}
\eeq
where $A_{ij}$ is the connection of a compact U(1) gauge field. The action of the U(1) gauge field is the standard generalization of the
Maxwell action
\beq
\mathcal{L}_A = K \sum_{\square}\cos \left( \sum_{\square} A_{ij} \right)  + \frac{1}{2h} \sum_{\langle ij \rangle} \left( \partial_\tau A_{ij} \right)^2 .
\label{SEA}
\eeq
For $p=0$ and large $\lambda$, we obtain the insulating N\'eel and
valence bond solid states \cite{NRSS89,NRSS90,FSX11}.
For $p\neq 0$ and large $\lambda$, 
the deconfined U(1)-ACL phase can appear at large $K$, while
the conventional phases A and B in 
Fig.~\ref{fig:higgs} appear at small $K$. We note, however, that the U(1)-ACL is expected to be unstable to pairing
and confinement, as was the case for the SU(2)-ACL \cite{MMSS14}.

\section{Conclusions}

We have reviewed candidate theories for describing the unconventional metallic phases observed over a wide region in the phase diagram of the cuprate high temperature superconductors. 

The key idea in the discussion has been to encapsulate the strongly correlated nature of the problem in terms of emergent gauge theories and topological order. In metals with well-defined quasiparticle excitations, we have shown how Luttinger's theorem allows us to sharply distinguish between phases with and without topological order. Specifically, we described a connection between the size
of the Fermi surface and the odd/even nature of the `symmetry enriched' \cite{WenSPT} TQFT describing the $\mathbb{Z}_2$ topological order.

A central mystery in the study of cuprate superconductors concerns the nature of the strange metal without quasiparticle excitations and its relation to an underlying quantum critical point. We have argued that the critical point is best described in terms of a transition between a metal with topological order and small Fermi surfaces, to a confining Fermi liquid with a large Fermi surface. Such a transition necessarily falls outside of the conventional Landau-Ginzburg-Wilson paradigm of symmetry-breaking phase transitions. 
Starting from a lattice model of electrons coupled to strongly fluctuating antiferromagnetic spin fluctuations, we proposed a deconfined critical theory for the strange metal, with a SU(2) gauge field coupled to a large Fermi surface of chargons and a critical Higgs field. 
On the low doping side of the critical point, the Higgs field condenses to leave a residual odd $\mathbb{Z}_2$ gauge theory
describing the pseudogap metal with small Fermi surfaces and long-range Ising-nematic order, and the magnitude of the pseudogap
determined by the magnitude of the Higgs field.
On the high doping side, the Higgs correlations are short-ranged, and 
the confining phase of the SU(2) gauge field leads to a large Fermi surface with no Ising-nematic order. 
Although the long-range Ising-nematic order vanishes at the critical point, the critical theory is not simply that of the onset
of this order in a Fermi liquid.

An overall perspective is provided by the phase diagram in Fig.~\ref{fig:higgs}.
This shows how the conventional physics of Hertz criticality, applicable to the pnictides, evolves 
smoothly to the `topological' physics of Higgs criticality, applicable to the hole-doped cuprates: the Higgs
field theory can be understood as a `SU(2) gauged' version of the Hertz theory. (Another perspective
on the phases in Fig.~\ref{fig:higgs} appears in a separate paper \cite{SBCS16}).
The Fermi surface reconstruction of electrons across the Hertz transition (from phase B to A)
is identical, at the saddle point level, to the Fermi surface reconstruction of chargons across the Higgs transition
(from phase D to phase C): this follows from the similarity between (\ref{model}) and (\ref{yukawac}). 
Consequently, the evolution of the charge transport across the conventional
transition between A and B will be similar to that across the topological transition between C and D.
Specifically, the evolution of the Hall effect as a function of $p$ in a model of the reconstruction of the Fermi surface
by long-range antiferromagnetism \cite{Storey16} also applies to the evolution of the Hall effect from C to D.
The remarkable agreement of such a model with recent observations \cite{LTCP15} makes the topological Higgs theory
an attractive candidate for the optimally hole-doped cuprates. Note that the use of fermionic chargons is important in this theory, 
and we argued that other approaches involving bosonic chargons \cite{LeeWenRMP,FG04} lead to rather different 
results for the Hall effect.

Future experiments will no doubt explore more completely the nature of the low $T$, high field, pseudogap metal
discovered in Ref.~\cite{LTCP15}. Quantum oscillations could yield more precise information on the nature of the Fermi surface,
including whether the quasiparticles are spinful (as in FL*) or spinless (as in ACL). Nuclear or muon spin resonance experiments
can determine if there is any field-induced magnetic order \cite{DSZ01} in this regime.
Also, studies of transport in the vicinity of the critical point between the pseudogap metal and the Fermi liquid, 
with techniques borrowed from hydrodynamics and holography, are promising avenues to explore. 

\subsection*{Acknowledgements}

We would like to especially thank Seamus Davis and Louis Taillefer for numerous discussions on experiment and theory which
have strongly influenced the perspective presented here.
The research was supported by the NSF under Grant DMR-1360789 and MURI grant W911NF-14-1-0003 from ARO.
Research at Perimeter Institute is supported by the Government of Canada through Industry Canada and by the Province of Ontario through the Ministry of Research and Innovation. SS also acknowledges support from Cenovus Energy at Perimeter Institute.

\bibliography{nambu}

\end{document}